\documentclass[a4paper,10pt,dvipsnames]{article}

\usepackage[utf8]{inputenc}
\usepackage{amsthm}
\usepackage{amsfonts}
\usepackage{amssymb}	
\usepackage{amsmath}
\usepackage{bbold}
\allowdisplaybreaks
\usepackage{mathtools}
\usepackage[english]{babel}
\usepackage{color}
\usepackage{slashed}
\usepackage{enumerate}
\usepackage{graphicx}
\usepackage{systeme}
\usepackage{dsfont}
\usepackage{relsize}
\usepackage[margin=0.90in]{geometry}
\usepackage{float}
\usepackage{soul}
\usepackage{tikz}
\usepackage[labelformat=simple]{subcaption}

\usepackage{graphicx}
\usepackage{bmpsize}
\usepackage{epstopdf}
\usepackage{bbm}
\usepackage{xcolor,etoolbox}
\usepackage{titling}
\thanksmarkseries{arabic} 
\usepackage{authblk}
\newcommand*\samethanks[1][\value{footnote}]{\footnotemark[#1]}
\usepackage{blindtext,graphicx}
\usepackage[absolute]{textpos}
\usepackage{physics}
\usepackage[hang,flushmargin]{footmisc}

\usepackage{xfrac}
\usepackage{nicefrac}
\usepackage{esvect}
\usepackage{cases}
\usepackage{empheq}
\usepackage{stmaryrd}
\usepackage{cancel}
\usepackage[linguistics]{forest}
\usepackage{url}
\usepackage[font=small]{caption}
\usepackage{multirow}
\usepackage{comment}
\newcommand{\fig}[1]{Fig. \ref{fig:#1}} 
\newcommand{\rrmm}{reduced relaxed micromorphic model } 
\newcommand{\rrm}{reduced relaxed micromorphic } 

\usepackage[most]{tcolorbox}
\usepackage{dashrule}
\usepackage{pifont}
\usepackage{fancybox}
\usepackage{fontawesome}
\usepackage[giveninits=true,sortcites=true,date=year,maxbibnames=99,doi=false,isbn=false,url=false,eprint=false]{biblatex}
\renewbibmacro{in:}{}
\DeclareFieldFormat{pages}{#1}
\AtEveryBibitem{%
  \clearlist{language}%
}
\usepackage{csquotes}

\usepackage{setspace}
\usepackage{subfiles} 

\title{Effective surface forces and non-coherent interfaces within the reduced relaxed micromorphic modeling of finite-size mechanical metamaterials}
\author{
Leonardo A. Perez Ramirez\thanks{leonardo.perez@tu-dortmund.de, +49 231 755 7262, Faculty of Architecture and Civil Engineering, TU Dortmund, August-Schmidt-Str. 8, Dortmund, Germany},
\quad
Félix Erel-Demore\thanks{Faculty of Architecture and Civil Engineering, TU Dortmund, August-Schmidt-Str. 8, Dortmund, Germany},
\quad
Gianluca Rizzi\samethanks[2],
\quad
Jendrik Voss\samethanks[2],
\quad
Angela Madeo\thanks{Head of Chair of Continuum Mechanics, Faculty of Architecture and Civil Engineering, TU Dortmund, \\ \indent \,\,\,\, August-Schmidt-Str. 8, Dortmund, Germany}
}
\begin{document}

\maketitle

\abstract{
This paper introduces for the first time the concepts of non-coherent interfaces and microstructure-driven interface forces in the framework of micromorphic elasticity.  It is shown that such concepts are of paramount importance when studying the response of finite-size mechanical metamaterials at the homogenized macro-scale. The need of introducing interface forces is elucidated through numerical examples comparing \rrm simulations to their full-microstructured  counterparts. These results provide a milestone for the understanding of metamaterials' modeling at the homogenized scale and for the use of micromorphic-type models to achieve an accurate upscaling towards larger-scale metamaterials' structures.
}


\pagebreak

\section{Introduction}
\label{sec:rmm_1}

As it is well known, surfaces of bodies and interfaces between pairs of bodies exhibit properties quite different from those associated with their bulk, due to the fact that molecules or other small-scale heterogeneities result to have different arrangements in proximity of the material’s boundaries than when they are forming a 3D network embedded in the material’s bulk \cite{Adam.1941,Adamson.1967,Gurtin.1975}.
This statement becomes even more true when considering so-called metamaterials, since the heterogeneities that come into play have characteristic sizes of many orders of magnitude bigger than typical molecules’ sizes and thus have greater chances to affect the macroscopic material behavior (both in the bulk and in proximity of interfaces) at those large scales that are familiar to engineers.

Metamaterials (from the Greek “meta” = beyond) are materials with architectured microstructures in which the mechanical properties of the microscopic components (unit cells) have an important effect at the macroscopic scale.
These metamaterials are often engineered by means of the periodic repetition in space of so-called unit cells that are designed to show special deformation mechanisms triggering exotic mechanical responses at the macroscopic scale.
In other words, the mechanical and geometrical properties of metamaterials’ microstructure are designed so that they strongly affect the overall bulk metamaterial properties.
Typical examples are metamaterials exhibiting band-gaps (frequency ranges in which elastic waves cannot propagate) \cite{Bilal.2018,Celli.2019,ElSherbiny.2018,Goh.2019,Koutsianitis.2019,Liu.2000,Wang.2014,Zhu.2015,Fedele.2023}, cloaking (elastic waves proceed unperturbed even if hitting the metamaterial) \cite{Buckmann.2015,Misseroni.2019,Misseroni.2016,Norris.2014,Rossi.2020}, focusing (a diffused incident wave is focused in a ray while passing inside the metamaterial) \cite{Cummer.2016,Guenneau.2007}, channeling (elastic waves take patterns with specific orientations while passing into the metamaterial) \cite{Bordiga.2019,Kaina.2017,Miniaci.2019,Tallarico.2017,Wang.2018}, negative refraction (waves are reflected in unusual ways when hitting an interface) \cite{Bordiga.2019,Lustig.2019,Morini.2019,Srivastava.2016,Willis.2016,Zhu.2015}, and many others, both in the static and dynamic regime.

In the last decades, many homogenization techniques have emerged trying to establish how to derive suitable macroscopic PDEs for mechanical metamaterials starting from specific microscopic unit-cells \cite{Allaire.1992,Andrianov.2008,Auriault.2012,Bensoussan.2011,Boutin.2014,Chen.2001,Craster.2010,Marigo.2016,Touboul.2020,Willis.2011,Willis.2012}.
However, since the target macroscopic model is generally chosen a priori to depend only on the displacement field (classical elasticity), the associated parameters (mass and/or stiffness) turn out to be frequency-dependent and may become negative for frequencies approaching the resonance frequency of the internal mass.

Recently, so-called computational homogenization techniques have been proposed that complement these upscaling techniques to include the possibility of letting enriched continua of the micromorphic type emerge at the macroscopic scale \cite{Liu.2021,Sridhar.2016}.
In the last 50 years countless efforts have been deployed to develop reliable effective models for the description of heterogeneous materials.
This has mainly been done in view of the tremendous advantage that a reliable effective model would bring in terms of new possibilities for meta-structural design at the large scales which are relevant for engineers (it is well known that fully microstructured simulations become unaffordable in terms of computational costs already for relatively few unit cells).
However, only very few authors seem to be aware of the crucial role that interface forces must play to develop realistic effective models for metamaterials.

Even if it is quite evident that interface at metamaterials’ boundaries must play a predominant role in the mechanical response of finite-size metamaterials, this issue is often disregarded, due to the complexity of adapting classical homogenization techniques to media with boundaries.
Indeed, using classical homogenization methods in realistic bounded domains is an issue which is an open scientific challenge \cite{Cornaggia.2023}.
Oscillating layers appearing near to interfaces can be accounted for in homogenization techniques by using corrector functions \cite{Moskow.1997,Josien.2019,Cakon.2016,Cakoni.2019,Vinoles.2019}.
However, these boundary correctors are complex objects that make difficult a thorough theoretical study \cite{Armstrong.2017,GerardVaret.2012} and their direct computation is often as costly as the one of the original problem defined on the microstructured domain \cite{Cornaggia.2023}.
Few works address the « practical implementation » of these correctors in view of possible numerical implementation and simulation \cite{Beneteau.2021,Vinoles.2019,Vinoles.2016,Marigo.2017,Maurel.2018}.
The time domain simulation of such homogenized models with boundaries is even more challenging and almost inexistent except for few noticeable examples \cite{Cornaggia.2023,Beneteau.2021}.
To deal with such complexity, in the present work, we adopt a different perspective with respect to bottom-up homogenization techniques and we postulate our problem directly at the macro-scale.
Leveraging the reduced relaxed micromorphic model that largely showed its performances for the description of metamaterials’ bulk propagation problems at the macro-scale \cite{Demore.2022,Demetriou.2023}, we complement it with the possibility of macroscopic traction discontinuities at the considered macroscopic boundaries, which take the form of « interface forces ».
Refraining from trying to obtain a general expression of these interface forces via a bottom-up approach (given the aforementioned difficulties), we explore the nature of such macroscopic interface forces based on a phenomenological approach.
In other words, we postulate the expression of such macroscopic interface forces based on \textit{i)} a comparison with the corresponding microstructured force and \textit{ii)} a comparison with the effect of such interface force on the reduced relaxed micromorphic solution for the displacement field.
We show that our procedure is able to bring new answers concerning the numerical implementation and simulation of realistic bounded metamaterials’ domains and that it open new perspectives to inspire the target boundary forces that should be looked for by bottom-up homogenization approaches.

It is intuitively clear that if the metamaterials’ micro-heterogeneities are so pronounced to have a tangible effect on the macroscopic metamaterials’ bulk response, they will have even more weight when considering macroscopic metamaterials’ interfaces that naturally occur when considering finite-size problems instead of infinite size domains.
It can be stated that such interface effects will play a greater role when decreasing the size of the considered macroscopic specimen.

It has been largely shown that the bulk mechanical response of mechanical metamaterials can be effectively explored using an elastic- and inertia-augmented micromorphic model (reduced relaxed micromorphic model) which is able to describe the main metamaterials’ fingerprint characteristics (anisotropy, dispersion, band-gaps, size-effects, etc.), while keeping a reduced structure (free of unnecessary parameters) \cite{Aivaliotis.2020,Madeo.2015,Neff.2020,Neff.2015,Neff.2014,Voss.2023,PerezRamirez.,Demetriou.2023,Rizzi.2021,Demore.2022,Rizzi.2022,Rizzi.2022b,Rizzi.2022c}.
This model can be linked a posteriori to real metamaterials’ microstructures via an inverse fitting procedure.
The reduced model’s structure, coupled with the introduction of well-posed boundary conditions, allowed us to unveil the dynamic response of metamaterials’ bricks of finite-size and complex shapes when microstructure-driven interface effects can be considered to be negligible.
Indeed, depending on the type of applied load and on the specific micro- and macro-geometry of the problem at hand, interface forces may still have a moderate effect on the overall metamaterial’s behavior \cite{Demetriou.2023}.
However, when considering more complex loading conditions, different unit cells and/or samples of small dimensions, interface effects cannot longer be ignored and the reduced relaxed micromorphic model must be enhanced to account for these effects.

In the present paper, we address for the first time, the fundamental issue of including interface effects in the reduced relaxed micromorphic framework by generalizing the typical interface models that can be found in the literature for classical elasticity.

Classical interface models can be divided into different classes based on continuity of certain variables associated with the problem at hand, namely the displacement and the traction at the considered interface.
In particular, the most common interface models can be briefly reviewed as follows:

\begin{itemize}
    \item \textit{Perfect interfaces.} Both the displacement and the traction are continuous across a perfect interface \cite{Javili.2017,Spannraft.2022,Firooz.2021,PerezRamirez.,Demetriou.2023,Rizzi.2021,Demore.2022,Rizzi.2022,Rizzi.2022b,Rizzi.2022c}.
    \item \textit{Cohesive interfaces.} If either the displacement or the traction has a jump across the interface, the interface model is imperfect. A cohesive interface is a particular imperfect interface and allows for displacement jumps across the interface but continuity of the traction is still satisfied \cite{Barenblatt.1959,Barenblatt.1962,Dugdale.1960,Needleman.1987,Xu.1994,Ortiz.1999,Charlotte.2006,Alfano.2001,Gasser.2003,vandenBosch.2006,Fagerstrom.2006,Park.2009,Mosler.2011,Park.2011,Dimitri.2015}.
    \item \textit{Elastic interfaces.} Contrarily to the cohesive interface model, the displacement is continuous across an elastic interface but the traction is discontinuous \cite{Murdoch.1976,Moeckel.1975,dellIsola.1987,Fried.2007,Gurtin.1975}.
    \item \textit{General interfaces.} Those interfaces allow both jumps in the displacement and the traction \cite{Javili.2017,Bovik.1994,Hashin.2002,Benveniste.2001,Monchiet.2010,Benveniste.2013}.
\end{itemize}

In the present paper we will only focus on the generalization of the concept of elastic interfaces in the context of micromorphic elasticity, because the considered metamaterials’ interfaces are always such that the macroscopic displacement remains continuous due to fast bonding of the solid phases across the interface.
However, the presence of voids inside the unit cells implies that the considered interfaces may also contain empty spaces on one or both sides, so that the traction can be expected to suffer jumps across these interfaces.

While the concept of « interface models » is well known in classical elasticity interface problems, their use is always limited to treat interfaces induced by damage \cite{Spannraft.2022,Barenblatt.1959,Barenblatt.1962,Dugdale.1960}, adhesives between two different media \cite{Spannraft.2022,Klarbring.1991,Klarbring.1998,Geymonat.1999,Xu.2003,Zhang.2009,Alfano.2009,Mubashar.2011,Campilho.2013}, surface effects in nano-systems \cite{Sharma.2004,Sharma.2007,Sharma.2003}, fracture \cite{Shih.1987,Tvergaard.1993,Wei.1999,Park.2009,Chandra.2002,Li.2005,Sun.2006,Park.2011}, delamination \cite{Espinosa.2000,Hu.2008,Aymerich.2008,Liu.2013,Parrinello.2016,Reinoso.2017}, crack growth \cite{England.1966,Tvergaard.1996,Yang.2001,Roe.2003,Bouvard.2009,Kawashita.2012}, bond failure \cite{Ingraffea.1984,Huang.2001}, screw dislocations \cite{Fan.2003}, grain boundaries \cite{Mori.1987,Pezzotta.2008,Wulfinghoff.2017}, peeling \cite{Wei.1998}.
No applications to mechanical metamaterials can be found until today.

When more theoretical studies are considered, the concept of «interface models» is almost always applied to infinite-size periodic structures obtained by the repetition in space of specific unit cells.
Those interfaces are always assumed to occur at the microscopic scale, namely, at the interface between two phases inside the unit cell \cite{Firooz.2021,Needleman.1987,Mogilevskaya.2008,Chatzigeorgiou.2017,Alfano.2001,Shuttleworth.1950,Sharma.2003,Dingreville.2005,Duan.2005,Benveniste.2001,Hashin.2002,Chen.2006,Javili.2013}: homogenization techniques are then applied to show that these non-perfect interfaces at the micro-level produce scale effects (elastic properties of the homogenized composite that depend on the size of the base unit cell) \cite{Javili.2017,Ottosen.2016,Firooz.2021,Javili.2015,Ban.2020,Yang.2004,Yvonnet.2008,Dingreville.2014,Chen.2020,Sharma.2004b,Zemlyanova.2018,Le.2020}.
To the authors’ knowledge, there is no study that considers generalized interfaces to model macroscopic metamaterials’ boundaries, interfaces between different metamaterials or between metamaterials and classical homogeneous materials.
When we talk about macroscopic interfaces, we mean that we consider a finite block of a given metamaterial which can be connected to another metamaterial block or to a block of homogeneous material: those interfaces can then be seen as composed by a finite number of unit cells which confer specific elastic properties to the interface itself.
Macroscopic interface forces can be activated at those macroscopic interfaces, due to the heterogeneity of the underlying microstructure at the lower scales.
The present paper will address this issue for the first time and will open new perspectives for future studies.

The paper is organized as follows:

In section \ref{sec:rmm_1} we address the fundamental issue of ``interface forces" occurring in the continuum modeling of mechanical metamaterials and we frame it in the context of the existing state of the art.
Section \ref{sec:rmm_2} is devoted to the generalization of the concept of ``elastic interface" in the context of \rrm elasticity.
In particular, we show how to deal with these surface forces occurring at \textit{i)} free micromorphic boundaries as a result of the presence of an underlying microstructure and \textit{ii)} at interfaces separating two \rrm media or a \rrm medium from a homogeneous material.
In section \ref{sec:rmm_3}, we show how to use the novel concept of \rrm interface forces to model free metamaterials' interfaces, and interfaces arising between two metamaterials or a metamaterial and a homogeneous material.
In section \ref{sec:rmm_4} we start unveiling the efficacy of the concept of \rrm interface forces with a specific example in which two different metamaterial's interfaces are generated by using two different ``unit cell's cuts" stemming from the same bulk metamaterial.
In section \ref{sec:rmm_5} we further explore the efficacy of the new continuum modeling framework through a more complex numerical example involving interfaces between a metamaterial and a homogeneous material.
Section \ref{sec:rmm_6} finally provides the conclusions about our findings and the multiple perspectives which are now open to drive the theoretical investigations about surface forces in the framework of \rrm elasticity towards tangible large-scale applications involving finite-size metamaterials as base building blocks.

\section{Equations of motion and boundary conditions for the reduced relaxed micromorphic model}
\label{sec:rmm_2}

\subsection{Bulk equations and ``free interface" boundary conditions}
\label{subsec:rmm_2_1}

The kinetic energy $K$ and strain energy $W$ of the \rrmm are shown in equations \eqref{eq:kinEneMic}-\eqref{eq:strEneMic} \cite{Demetriou.2023,Rizzi.2021,PerezRamirez.,Voss.2023}:

\begin{align}
K \left(\dot{u},\nabla \dot{u}, \dot{P}\right) 
=&
\dfrac{1}{2}\rho \, \langle \dot{u},\dot{u} \rangle + 
\dfrac{1}{2} \langle \mathbb{J}_{\rm m}  \, \text{sym} \, \dot{P}, \text{sym} \, \dot{P} \rangle 
+ \dfrac{1}{2} \langle \mathbb{J}_{\rm c} \, \text{skew} \, \dot{P}, \text{skew} \, \dot{P} \rangle
\notag
\\
&
+ \dfrac{1}{2} \langle \mathbb{T}_{\rm e} \, \text{sym}\nabla \dot{u}, \text{sym}\nabla \dot{u} \rangle
+ \dfrac{1}{2} \langle \mathbb{T}_{\rm c} \, \text{skew}\nabla \dot{u}, \text{skew}\nabla \dot{u} \rangle
\label{eq:kinEneMic}
\ ,
\\
W \left(\nabla u, P\right)
=& 
\dfrac{1}{2} \langle \mathbb{C}_{\rm e} \, \text{sym}\left(\nabla u -  \, P \right), \text{sym}\left(\nabla u -  \, P \right) \rangle
\notag
\\
&
+ \dfrac{1}{2} \langle \mathbb{C}_{\rm c} \, \text{skew}\left(\nabla u -  \, P \right), \text{skew}\left(\nabla u -  \, P \right) \rangle
\notag
\\
&
+ \dfrac{1}{2} \langle \mathbb{C}_{\rm micro} \, \text{sym}  \, P,\text{sym}  \, P \rangle
\label{eq:strEneMic} 
\ .
\end{align}

The Lagrangian density can thus be written as:

\begin{align}
\mathcal{L}
\coloneqq&
K \left(\dot{u},\nabla \dot{u}, \dot{P}\right) 
-
W \left(\nabla u, P\right)
\ .
\label{eq:lagMicrom} 
\end{align}

In the previous formulas, $u \in \mathbb{R}^{3}$ is the macroscopic displacement field, $P \in \mathbb{R}^{3\times 3}$ is the non-symmetric micro-distortion tensor, $\rho$ is the macroscopic apparent density, $\mathbb{J}_{\rm m}$, $\mathbb{J}_{\rm c}$, $\mathbb{T}_{\rm e}$, $\mathbb{T}_{\rm c}$ are 4th order micro-inertia tensors, and $\mathbb{C}_{\rm e}$, $\mathbb{C}_{\rm m}$, $\mathbb{C}_{\rm c}$ are 4th order elasticity tensors.
These tensors in Voigt notation, for the tetragonal symmetry case and reporting only the in-plane components, can be expressed as \cite{Aivaliotis.2020,dAgostino.2020}:\footnote{Where in \cite{Aivaliotis.2020,dAgostino.2020} $\kappa_i = \lambda_i + \mu_i $ with $i =\{e,m\}$, $\kappa_{\gamma} = \gamma_3 + \gamma_1 $, $\overline{\kappa}_{\gamma} = \overline\gamma_3 + \overline\gamma_1 $, and $\eta_i = \rho L_c^2 \gamma_i$.}

\begin{align}
        \mathbb{C}_{\rm e}
        &= 
        \begin{pmatrix}
        \kappa_{\rm e} + \mu_{\rm e}	& \kappa_{\rm e} - \mu_{\rm e}				& \star & \dots	& 0\\ 
        \kappa_{\rm e} - \mu_{\rm e}	& \kappa_{\rm e} + \mu_{\rm e} & \star & \dots & 0\\
        \star & \star & \star & \dots & 0\\
        \vdots & \vdots	& \vdots & \ddots &\\ 
        0 & 0 & 0 & & \mu_{\rm e}^{*}
        \end{pmatrix},
        &
        \mathbb{C}_{\rm micro}
        &=
        \begin{pmatrix}
        \kappa_{\rm m} + \mu_{\rm m}	& \kappa_{\rm m} - \mu_{\rm m}				& \star & \dots	& 0\\ 
        \kappa_{\rm m} - \mu_{\rm m}	& \kappa_{\rm m} + \mu_{\rm m} & \star & \dots & 0\\
        \star & \star & \star & \dots & 0\\
        \vdots & \vdots	& \vdots & \ddots &\\ 
        0 & 0 & 0 & & \mu_{\rm m}^{*}
        \end{pmatrix},
        \notag
        \\[2mm]
        \mathbb{J}_{\rm m}
        &=
        \rho L_{\rm c}^2
        \begin{pmatrix}
        \kappa_\gamma + \gamma_{1} & \kappa_\gamma - \gamma_{1} & \star & \dots & 0\\ 
        \kappa_\gamma - \gamma_{1} & \kappa_\gamma + \gamma_{1} & \star & \dots & 0\\ 
        \star & \star & \star & \dots & 0\\
        \vdots & \vdots & \vdots & \ddots &\\ 
        0 & 0 & 0 & & \gamma^{*}_{1}\\ 
        \end{pmatrix},
        &
        \mathbb{T}_{\rm e}
        &=
        \rho L_{\rm c}^2
        \begin{pmatrix}
        \overline{\kappa}_{\gamma} + \overline{\gamma}_{1} & \overline{\kappa}_{\gamma} - \overline{\gamma}_{1} & \star & \dots	& 0\\ 
        \overline{\kappa}_{\gamma} - \overline{\gamma}_{1} &   \overline{\kappa}_{\gamma} + \overline{\gamma}_{1} & \star & \dots & 0\\ 
        \star & \star & \star & \dots & 0\\
        \vdots & \vdots & \vdots & \ddots &\\ 
        0 & 0 & 0 & & \overline{\gamma}^{*}_{1}
        \end{pmatrix},
        \label{eq:tensors}
\end{align}

\begin{align}
        \mathbb{J}_{\rm c}
        &=
        \rho L_{\rm c}^2
        \begin{pmatrix}
        \star & 0 & 0\\ 
        0 & \star & 0\\ 
        0 & 0 & 4\,\gamma_{2}
        \end{pmatrix},
        &
        \mathbb{T}_{\rm c}
        &=
        \rho L_{\rm c}^2
        \begin{pmatrix}
        \star & 0 & 0\\ 
        0 & \star & 0\\ 
        0 & 0 & 4\,\overline{\gamma}_{2}
        \end{pmatrix},
        &
        \mathbb{C}_{\rm c}
        &= 
        \begin{pmatrix}
        \star & 0 & 0\\ 
        0 & \star & 0\\ 
        0 & 0 & 4\,\mu_{\rm c}
        \end{pmatrix}.
        \notag
\end{align}

The action functional $\mathcal{A}$ of the micromorphic continuum is defined by:

\begin{align}
\mathcal{A}=\iint\limits_{\Omega \times \left[0,T\right]} 
\mathcal{L} \left(\nabla u, \dot{u}, \nabla \dot{u}, P, \dot{P}\right)
\, dx \, dt \, ,
\label{eq:act_func}
\end{align}

\noindent
and it operates on a domain $\Omega$ over the time interval $\left[0,T\right]$.
If we consider a conservative system, the work of internal actions corresponds to $\mathcal{W}^{int} \coloneqq \partial \mathcal{A}$, here the variation operator $\delta$ operates on the kinematic fields $(u,P)$.
If no external forces act on the considered system, an equilibrium condition is found when the first variation of the action functional is equal to zero, $\partial \mathcal{A} = 0$.
This, in turn, implies the strong form equilibrium equations:

\begin{equation}
\rho\,\ddot{u} - \text{Div}\widehat{\sigma} = \text{Div}\widetilde{\sigma}
\qquad\qquad
\mathrm{and}
\qquad\qquad
\overline{\sigma} = \widetilde{\sigma} - s
\qquad\qquad
\mathrm{in} \quad
\Omega
\, ,
\label{eq:equiMic}
\end{equation}

\noindent
together with the boundary conditions:

\begin{tcolorbox}[colback=Salmon!15!white,colframe=Salmon!50!white,coltitle=black,title={\centering \textbf{Generalized traction / displacement boundary conditions\\at a \rrm free interface}}]
\begin{align}
	\underset{
    \mathrm{vanishing\ generalized\ traction}
    }
    {\raisebox{0.5\baselineskip}{$t \coloneqq \left(\widetilde{\sigma} + \widehat{\sigma} \right) \, n = 0$}}
	\qquad\qquad
    \mathrm{or}
    \qquad\qquad
    \underset{
	\mathrm{assigned\ displacement}
    }
    {\raisebox{0.5\baselineskip}{$u = \overline{u}$}}
    \qquad\qquad
    \mathrm{on} \quad
    \partial \Omega \, .
	\label{eq:tractions}
\end{align}
\vspace{0\baselineskip}
\end{tcolorbox}

In the previous formulas, we set:

\begin{align}
\widetilde{\sigma}
&
\coloneqq \mathbb{C}_{\rm e}\,\text{sym}(\nabla u-P) + \mathbb{C}_{\rm c}\,\text{skew}(\nabla u-P)
\, ,
&
\widehat{\sigma}
&
\coloneqq \mathbb{T}_{\rm e}\,\text{sym} \, \nabla\ddot{u} + \mathbb{T}_{\rm c}\,\text{skew} \, \nabla\ddot{u}
\,,
\notag
\\
s
&
\coloneqq \mathbb{C}_{\rm micro}\, \text{sym} P
\,,
&
\overline{\sigma}
&
\coloneqq \mathbb{J}_{\rm m}\,\text{sym} \, \ddot{P} + \mathbb{J}_{\rm c}\,\text{skew} \, \ddot{P}
\,.
\label{eq:equiSigAll}
\end{align}

It is possible to understand that when a metamaterial interface possesses an underlying microstructure, homogenized surface forces can be generated due to the microstructure itself, especially for some frequency ranges (e.g., band gap regions) and for specific loading conditions.
This leads us to modify the traction boundary condition \eqref{eq:tractions} as follows:\footnote{The modified form \eqref{eq:forces_interface} of the boundary conditions \eqref{eq:tractions} can be easily re-formulated in the framework of the principle of virtual works by adding a specific surface contribution involving $f_{\rm interface}$ to the work of external actions.}

\begin{tcolorbox}[colback=BlueGreen!15!white,colframe=BlueGreen!50!white,coltitle=black,title={\centering \textbf{Interface force / displacement boundary conditions\\at a \rrm free interface}}]
\begin{align}
    \underset{
    \mathrm{interface\ force}
    }
    {\raisebox{0.5\baselineskip}{$t \coloneqq \left(\widetilde{\sigma} + \widehat{\sigma} \right) \, n = f_{\rm interface}$}}
	\qquad\qquad
    \mathrm{or}
    \qquad\qquad
    \underset{
	\mathrm{assigned\ displacement}
    }
    {\raisebox{0.5\baselineskip}{$u = \overline{u}$}}
    \qquad\qquad
    \mathrm{on} \quad
    \partial \Omega \, ,
    \label{eq:forces_interface}
\end{align}
\vspace{0\baselineskip}
\end{tcolorbox}

\noindent
where $f_{\rm interface}$ is a microstructure-driven interface force that may depend on the frequency, on the type of applied load and on the microstructure's geometry close to the interface.
In particular, this interface force will depend on the type of unit cell's cut which is chosen for a given metamaterial (see e.g., \fig{alpha_beta_cuts_together} and \ref{fig:alpha_beta_cuts}).

Contrarily to equation \eqref{eq:tractions}$_1$, the interface force boundary conditions introduced in equation \eqref{eq:forces_interface}$_1$ allow us to discriminate between two different unit cell's cuts through the introduction of this interface force $f_{\rm interface}$, while remaining in the simplified \rrm framework (no need to specify the specific microstructures in a FEM implementation).
In fact if, on the one hand, the generalized traction $t$ allows us to account for some microstructure-related effects through the introduction of extra terms with respect to classical elasticity, on the other hand, these extra terms come from the metamaterial's bulk behavior encoded in the expression \eqref{eq:act_func} of the action functional.
This means that the generalized traction $t$ alone does not allow us to discriminate between two different unit cell's cuts, since the bulk metamaterial generated by the two cuts is the same (see \fig{alpha_beta_cuts_together}).

The difference between two unit cell's cuts giving rise to the same metamaterial becomes visible only in the vicinity of the macroscopic metamaterial's intefaces (see \fig{alpha_free_interface} and \ref{fig:beta_free_interface}): this motivated us to introduce the ``interface force" $f_{\rm interface}$ in equation \eqref{eq:forces_interface}$_1$, which may account for such interface effects.
Even if it is not possible to give \textit{``a priori"} a comprehensive expression for $f_{\rm interface}$, we can state in general that it will depend on \textit{i)} the type of unit cell's cut, \textit{ii)} the type of applied load, and \textit{iii)} the considered frequency.
We will provide explicit expressions for $f_{\rm interface}$ in section \ref{sec:rmm_4}, where a specific ``free interface" problem is addressed.

\subsection{Interface conditions for coherent and non-coherent \rrm interfaces}

The interface conditions between two reduced relaxed micromorphic domains $\Omega^-$ and $\Omega^+$, which are in contact through a surface $\Sigma$, in the absence of external forces, can be derived via the minimization of the action functional $\mathcal{A}$ of equation \eqref{eq:act_func}, and making use of test functions with compact support inlcuding portions of $\Sigma$ \cite{dellIsola.2009}:

\begin{tcolorbox}[colback=Salmon!15!white,colframe=Salmon!50!white,coltitle=black,title={\centering \textbf{Continuous traction and displacement boundary conditions\\at a \rrm / \rrm coherent interface}}]
\begin{align}
	\underset{
    \mathrm{continuity\ of\ traction}
    }
    {\raisebox{0.5\baselineskip}{$t^+ = t^-$}}
    \qquad\qquad
    \mathrm{and}
    \qquad\qquad
	\underset{
    \mathrm{continuity\ of\ displacement}
    }
    {\raisebox{0.5\baselineskip}{$u^+ = u^-$}}
    \, .
	\label{eq:interface_conditions_micromorphic}
\end{align}
\vspace{0\baselineskip}
\end{tcolorbox}

As a particular case, if $\Omega^+$ is a classical isotropic Cauchy continuum, then the interface conditions reduce to

\begin{tcolorbox}[colback=Salmon!15!white,colframe=Salmon!50!white,coltitle=black,title={\centering \textbf{Continuous traction and displacement boundary conditions\\at a Cauchy / \rrm coherent interface}}]
\begin{align}
	\underset{
    \mathrm{continuity\ of\ traction}
    }
    {\raisebox{0.5\baselineskip}{$t_{\rm Cauchy}^+ = t^-$}}
	\qquad\qquad
    \mathrm{and}
    \qquad\qquad
    \underset{
    \mathrm{continuity\ of\ displacement}
    }
    {\raisebox{0.5\baselineskip}{$u^+ = u^-$}}
    \, ,
	\label{eq:interface_conditions_cauchy}
\end{align}
\vspace{0\baselineskip}
\end{tcolorbox}

\noindent
where $t_{\rm Cauchy}=\sigma_{\rm Cauchy} n$, and $\sigma_{\rm Cauchy}=\lambda \, \text{tr}\left(\text{sym} \nabla u\right) \, \mathbb{1} + 2\mu \, \text{sym}\nabla u$ is the stress tensor for an isotropic linear elastic material.

Interface conditions for which tractions and displacements are both continuous across the interface are known in classical elasticity as ``coherent interfaces" (see \cite{Javili.2017,Spannraft.2022,Firooz.2021}).
We will keep this nomenclature in the generalized framework of micromorphic elasticity.

As already remarked before for the ``free interface" conditions, the continuity of generalized tractions given in equation \eqref{eq:interface_conditions_micromorphic}$_{1}$ (or \eqref{eq:interface_conditions_cauchy}$_{1}$) accounts for the fact that the considered (macro-)material possesses an underlying microstructure via the introduction of a generalized traction force containing additional terms with respect to classical Cauchy elasticity (see definitions of $\widetilde{\sigma}$ and $\widehat{\sigma}$ in equation \eqref{eq:equiSigAll}).
However, these additional terms only account for those microstructure-related effects which come from the bulk and do not allow to discriminate between two different unit cell's cuts in the vicinity of interfaces.
With an analogous reasoning as the one drawn in section \ref{subsec:rmm_2_1}, we modify equations \eqref{eq:interface_conditions_micromorphic} and \eqref{eq:interface_conditions_cauchy} as follows, in order to account for the presence of interface forces directly driven by different unit cell's cuts:

\begin{tcolorbox}[colback=BlueGreen!15!white,colframe=BlueGreen!50!white,coltitle=black,title={\centering \textbf{Reduced relaxed micromorphic / \rrm elastic (non-coherent) interface}}]
\begin{align}
    \underset{
    \mathrm{jump\ of\ traction}
    }
    {\raisebox{0.5\baselineskip}{$\llbracket t \rrbracket \coloneqq t^+ - t^- = f \neq 0$}}
    \quad
    \mathrm{and}
    \quad
	\underset{
    \mathrm{continuity\ of\ displacement}
    }
    {\raisebox{0.5\baselineskip}{$u^+ = u^-$}}
    \quad
    ,
    \quad
    \begin{tabular}{cc}
         $f:$ & $\partial \Omega \times \mathbb{R}_+ \to \mathbb{R}^3$\\
             & $x,\omega \mapsto f \left( x, \omega \right)$
    \end{tabular}
    \, ,
	\label{eq:interface_non_coherent_conditions_micromorphic}
\end{align}
\vspace{0\baselineskip}
\end{tcolorbox}

\noindent
and

\begin{tcolorbox}[colback=BlueGreen!15!white,colframe=BlueGreen!50!white,coltitle=black,title={\centering \textbf{Cauchy / \rrm elastic (non-coherent) interface}}]
\begin{align}
    \underset{
    \mathrm{jump\ of\ traction}
    }
    {\raisebox{0.5\baselineskip}{$\llbracket t \rrbracket \coloneqq t_{\rm Cauchy}^+ - t^- = f \neq 0$}}
    \quad
    \mathrm{and}
    \quad
	\underset{
    \mathrm{continuity\ of\ displacement}
    }
    {\raisebox{0.5\baselineskip}{$u^+ = u^-$}}
    \quad
    ,
    \quad
    \begin{tabular}{cc}
         $f:$ & $\partial \Omega \times \mathbb{R}_+ \to \mathbb{R}^3$\\
             & $x,\omega \mapsto f \left( x, \omega \right)$
    \end{tabular}
    \, .
	\label{eq:interface_non_coherent_conditions_cauchy}
\end{align}
\vspace{0\baselineskip}
\end{tcolorbox}

This modification of the micromorphic traction boundary conditions generalizes what is known in the literature as ``boundary conditions for non-coherent interfaces" for the case of classical elasticity (see \cite{Javili.2017,Ottosen.2016}).
Non-coherent interfaces can be found between two materials that are non-homogeneous close to the considered interface due to the presence of voids or defects.
The different possible cases that are possible in a ``continuum mechanics" framework to describe non-coherent interfaces, are those allowing jumps of tractions but not displacement jumps (elastic interfaces), those allowing displacement jumps but not traction jumps (cohesive interfaces), and those allowing both traction and displacement jumps (general interfaces) (see \cite{Javili.2017} and section\ref{sec:rmm_1}).

Given the specific characteristics of the materials considered in this paper, it is clear that, since the two media are always strongly connected at the considered interfaces, the macroscopic displacement jump must be considered to be vanishing (continuity of macroscopic displacement).
On the other hand, given the presence of voids in the vicinity of the considered interfaces, the jump of traction can be non-vanishing.
In this work, we will thus limit ourselves to generalize only elastic (non-coherent) interface conditions in the framework of micromorphic elasticity.
It is worth to explicitly remark that imposing non-vanishing traction jumps turns out to be equivalent to the existence of a microstructure-related surface force as in the ``free interface" case considered in section \ref{subsec:rmm_2_1}.

\section{Interfaces in mechanical metamaterials}
\label{sec:rmm_3}

The elastic interface conditions newly introduced in equations \eqref{eq:forces_interface}-\eqref{eq:interface_conditions_cauchy} allow us to treat, for the first time in the literature, complex interface problems for finite-size metamaterials' structures, in the macroscopic framework of (reduced relaxed) micromorphic elasticity.
Indeed, while the concept of ``non-coherent interfaces" is well known in classical elasticity interface problems, it has never been used neither to explore complex macroscopic interfaces in mechanical metamaterials, nor to enhance the performances of micromorphic-type continuum models with respect to an effective description of metamaterials' interfaces (see section \ref{sec:rmm_1} for a thorough review of ``non-coherent interfaces" and their typical application fields).

\begin{figure}[H]
     \centering
     \begin{subfigure}[b]{\textwidth}
         \centering
         \includegraphics[width=\textwidth]{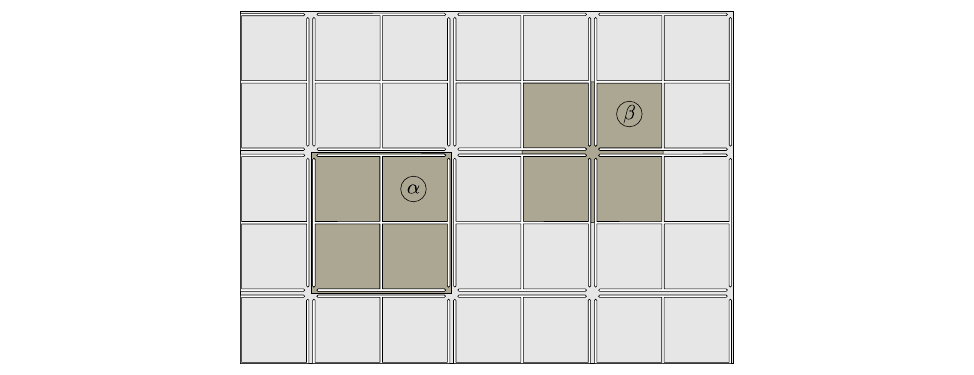}
     \end{subfigure}
        \caption{Two possible unit cell's cuts giving rise to the same metamaterial: $\alpha$ and $\beta$ cell's cuts.}
        \label{fig:alpha_beta_cuts_together}
\end{figure}

In order to provide a concrete example of how elastic interfaces should be implemented in a \rrm framework to describe complex problems arising at metamaterials' interfaces, we consider here a metamaterial stemming from the periodic repetition in space of the unit cell presented in \cite{Demore.2022} (see \fig{alpha_beta_cuts_together}).

\begin{figure}[H]
     \centering
     \begin{subfigure}[b]{\textwidth}
         \centering
         \includegraphics[width=\textwidth]{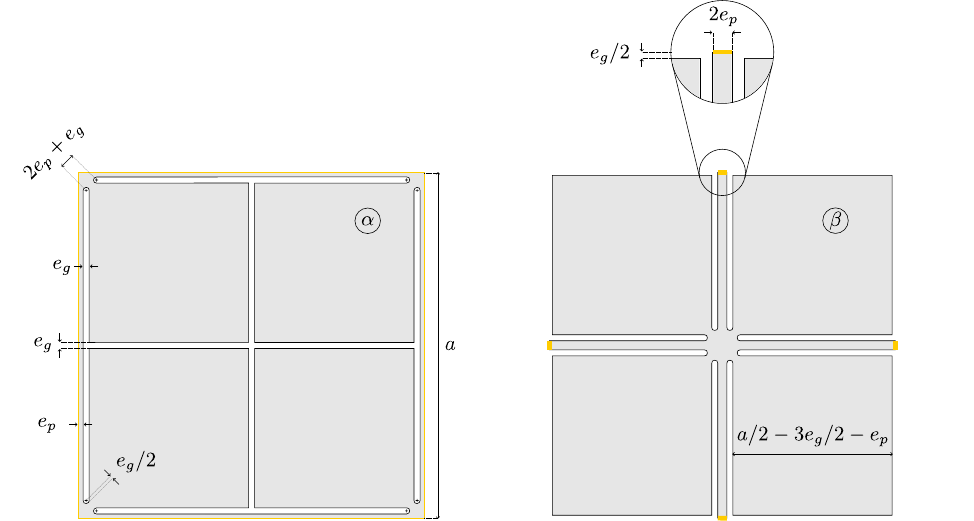}
     \end{subfigure}
        \caption{$\alpha$ and $\beta$ unit cell's cuts for the metamaterial in \fig{alpha_beta_cuts_together}. Solid connections are indicated in yellow.}
        \label{fig:alpha_beta_cuts}
\end{figure}

These unit cell's cuts $\alpha$ and $\beta$ are two possible extremes allowing to build the metamaterial in \fig{alpha_beta_cuts_together}, while keeping the tetragonal symmetry: unit cell $\alpha$ provides an interface which is everywhere solid (even if voids are close to the boundary), while unit cell $\beta$ provides an interface which is empty ``almost everywhere" except for small solid connections given by the four ``slender" beams (see \fig{alpha_beta_cuts}).

It is clear that many other different cell's cuts with tetragonal symmetry would be possible that give rise to the same bulk metamaterial.
However, those cell's cuts would provide interfaces which are intermediate cases between the $\alpha$ cut (full solid interface) and the $\beta$ cut (almost empty) interface.
The base material of the unit cells in \fig{alpha_beta_cuts} is titanium, its elastic and geometric properties are given in Table \ref{table:unit_cell_properties}.

\begin{table}[ht]
\centering
\begin{tabular}{||c c c c c c||} 
 \hline
 $a$ & $e_g$ & $e_p$ & $\rho_{\mathrm{Ti}}$ & $\lambda_{\mathrm{Ti}}$ & $\mu_{\mathrm{Ti}}$
 \\[1mm]
 [mm] & [mm] & [mm] & [kg/m$^3$] & [GPa] & [GPa]
 \\[1mm]
 \hline\hline
 20 & 0.35 & 0.25 & 4400 & 88.8 & 41.8
 \\
 \hline
\end{tabular}
\caption{Geometrical and elastic parameters of the metamaterial's unit cell.}
\label{table:unit_cell_properties}
\end{table}

\begin{figure}[H]
     \centering
     \begin{subfigure}[b]{\textwidth}
         \centering
         \includegraphics[width=\textwidth]{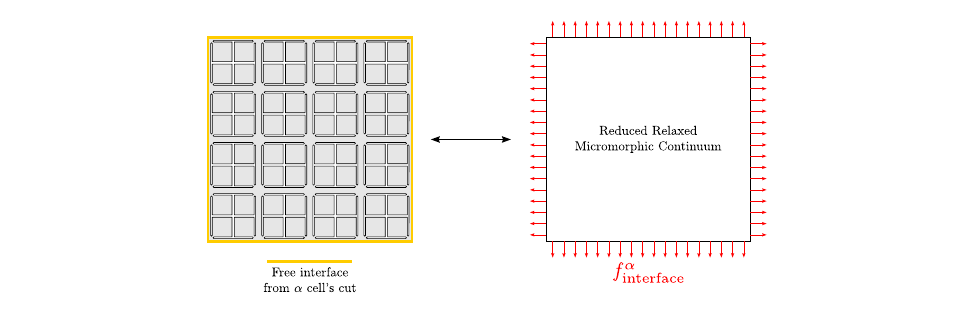}
     \end{subfigure}
        \caption{Free macroscopic metamaterial interface for a metamaterial block stemming from the $\alpha$ cell's cut \textit{(left)} and its reduced relaxed micromorphic counterpart, modeled via the introduction of the interface force $f_{\mathrm{interface}}^{\alpha}$ \textit{(right)}.}
        \label{fig:alpha_free_interface}
\end{figure}

We want to underline again that in the present paper we are interested in ``macroscopic interfaces" which arise as boundaries of finite-size metamaterial's blocks, and not in those micro-interfaces that would arise inside the unit cell if the voids would be filled by a different material.
Given the fact that the underlying unit cells contain voids, we suppose that the \rrm interfaces will be subjected to macroscopic interface forces (or jump of forces) directly stemming from the heterogenity of the unit cells which are adjacent to the considered interfaces (see \fig{alpha_free_interface}).

This is a quite different approach to ``non-coherent interfaces" with respect to the one usually found in the literature, where such generalized interfaces are always considered to arise at the microscopic level, as separation interfaces between two phases inside metamaterial's unit cells \cite{Javili.2017}.

\begin{figure}[H]
     \centering
     \begin{subfigure}[b]{\textwidth}
         \centering
         \includegraphics[width=\textwidth]{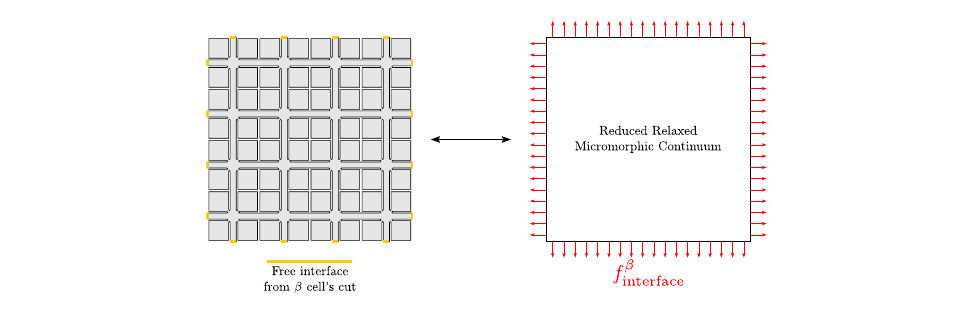}
     \end{subfigure}
        \caption{Free macroscopic metamaterial interface for a metamaterial block stemming from the $\beta$ cell's cut (left) and its reduced relaxed micromorphic counterpart, modeled via the introduction of the interface force $f_{\mathrm{interface}}^{\beta}$ (right).}
        \label{fig:beta_free_interface}
\end{figure}

When one of our macroscopic interfaces coincides with the boundary of a finite-size metamaterial block, we will call it a ``free" metamaterial interface (see \fig{alpha_free_interface} and \ref{fig:beta_free_interface}).

\begin{figure}[H]
     \centering
     \begin{subfigure}[b]{\textwidth}
         \centering
         \includegraphics[width=\textwidth]{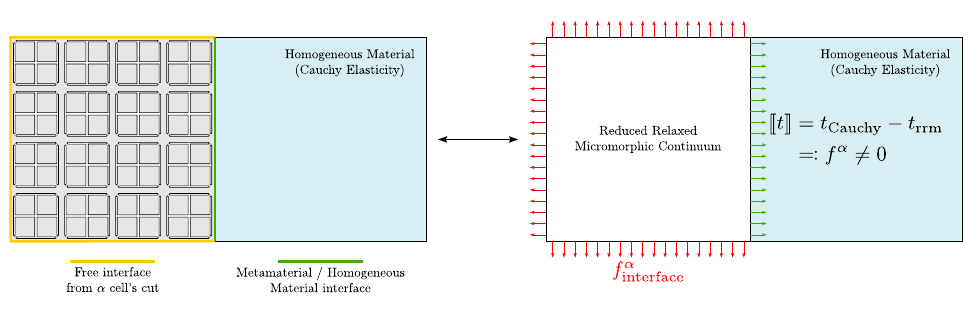}
     \end{subfigure}
        \caption{Metamaterial interface \textit{(green line)} between a metamaterial stemming from the unit cell cut $\alpha$ and a homogeneous material \textit{(left)} and its \rrm counterpart modeled through the introduction of a non-vanishing traction jump.}
        \label{fig:alpha_hom_interface}
\end{figure}

\begin{figure}[H]
     \centering
     \begin{subfigure}[b]{\textwidth}
         \centering
         \includegraphics[width=\textwidth]{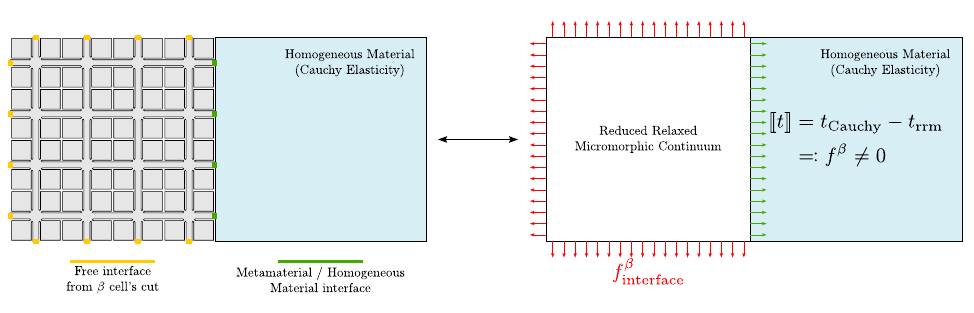}
     \end{subfigure}
        \caption{Metamaterial interface \textit{(green line)} between a metamaterial stemming from the unit cell cut $\beta$ and a homogeneous material \textit{(left)} and its \rrm counterpart modeled through the introduction of a non-vanishing traction jump.}
        \label{fig:beta_hom_interface}
\end{figure}

In the case in which the macroscopic interface separates two metamaterials or a metamaterial from a homogeneous material, we will just call it metamaterial interface (see \fig{alpha_hom_interface} and \ref{fig:beta_hom_interface}).

It is clear that different unit cell's cuts (see \fig{alpha_beta_cuts_together}) will give rise to different macroscopic metamaterial's interfaces and consequently to different interface forces $f_{\rm interface}$ when passing to the reduced relaxed modeling, see \fig{alpha_free_interface} and \ref{fig:beta_free_interface}.

A similar reasoning holds for metamaterial interfaces between two metamaterials or between a metamaterial and a homogeneous material, with the only difference that the interface force has the indirect effect of generating a traction jump across the considered interface.
We show in \fig{alpha_hom_interface} and \ref{fig:beta_hom_interface} the definition of metamaterial's interfaces between a metamaterial and a homogeneous material (the case of an interface between two metamaterials would be completely analogous). In these figures, the homogeneous material is an arbitrary material without voids that follows Cauchy Elasticity.

We have thus set in this section the general framework to treat ``free" metamaterial interfaces and metamaterial interfaces between pairs of media in the simplified framework of \rrm elasticity.
We will show in the folowing section that, depending on i) the unit cell cut, ii) the type of loading and/or the considered frequency, the effect of the interface forces may become negligible, so that one could simply use the classical \rrm modeling where $f_{\rm interface}=0$, which implies $\llbracket t \rrbracket=0$.
On the other hand, when considering unit cell's cuts which have bigger fractions of voids along the interface and/or specific frequencies (e.g., band-gap frequencies), the correct modeling with $f_{\rm interface} \neq 0$ and $\llbracket t \rrbracket \neq 0$ can no longer be ignored.
We will expand these findings in sections \ref{sec:rmm_4} and \ref{sec:rmm_5}, where specific interface problems are analyzed.

\section{Effect of different cell's cuts on a metamaterial's ``free interface" problem}
\label{sec:rmm_4}

\begin{figure}[H]
     \centering
     \begin{subfigure}[b]{\textwidth}
         \centering
         \includegraphics[width=\textwidth]{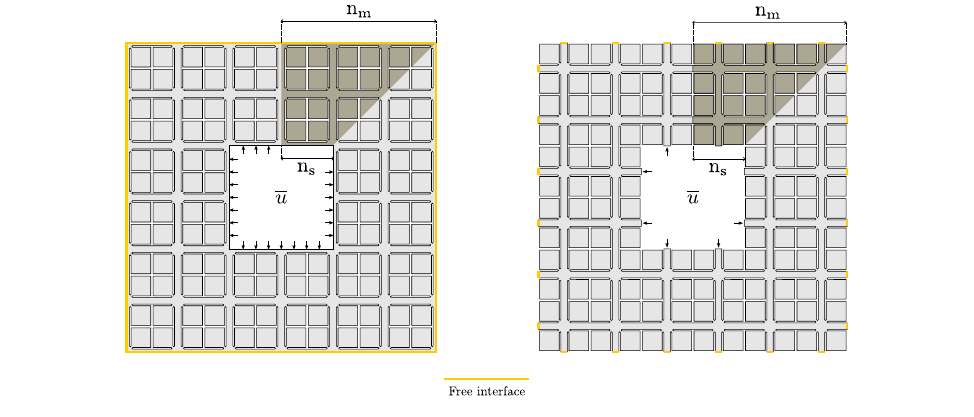}
     \end{subfigure}
        \caption{Full-microstructure setup of the free interface problem for \textit{(left)} $\alpha$ and \textit{(right)} $\beta$ cell's cuts.}
        \label{fig:geometry_free}
\end{figure}

We start by implementing in a finite element code, the ``free interface" problem defined in \fig{geometry_free}, where a metamaterial plate with a central hole is loaded by adding a harmonic radial expansion displacement inside the hole. We use the Structural Mechanics Module of COMSOL Multiphysics® software to perform the simulations \cite{Comsol}.  
Both the microstructured simulations ($\alpha$ and $\beta$ cuts) and the \rrm simulations are implemented in COMSOL Multiphysics®.
To implement the full microstructured simulations we implement the detailed geometries and we simply use the classical elasticity theory to model the mechanical response of the solid phase which is supposed to be Titanium (see Table \ref{table:unit_cell_properties}).
As for the implementation of the relaxed micromorphic simulations, we implement an analogous continuum domain with boundary conditions \eqref{eq:forces_interface}, as depicted in \fig{rrm_implementation_simple_plate}.

In both cases, to reduce computation time, we implement symmetry conditions simplifying the study to the problem in \fig{alpha_beta_free} (see Appendix \ref{subsec:rmm_7_1} for details about the implementation of the symmetry conditions).

\begin{figure}[H]
     \centering
     \begin{subfigure}[b]{\textwidth}
         \centering
         \includegraphics[width=\textwidth]{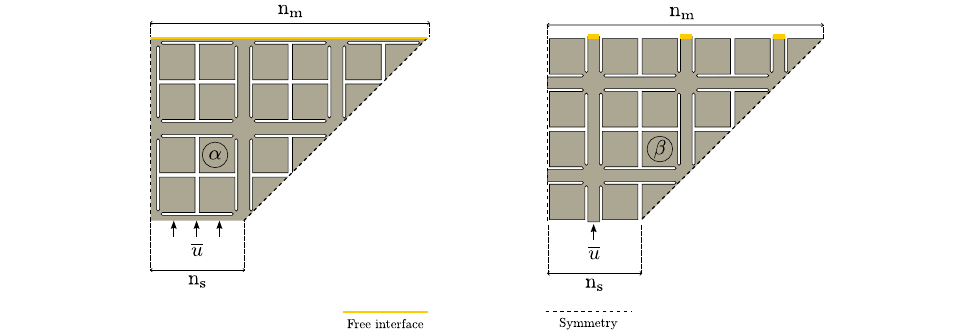}
     \end{subfigure}
        \caption{Symmetrized domain for the \textit{(left)} $\alpha$ and \textit{(right)} $\beta$ unit cells.}
        \label{fig:alpha_beta_free}
\end{figure}

As for the implementation of the relaxed micromorphic simulations, we implement an analogous continuum domain with boundary conditions \eqref{eq:forces_interface}, as depicted in \fig{rrm_implementation_simple_plate}.
Also, in the \rrm simulation we implement symmetry conditions as described in Appendix \ref{subsec:rmm_7_1}.

\begin{figure}[H]
     \centering
     \begin{subfigure}[b]{\textwidth}
         \centering
         \includegraphics[width=\textwidth]{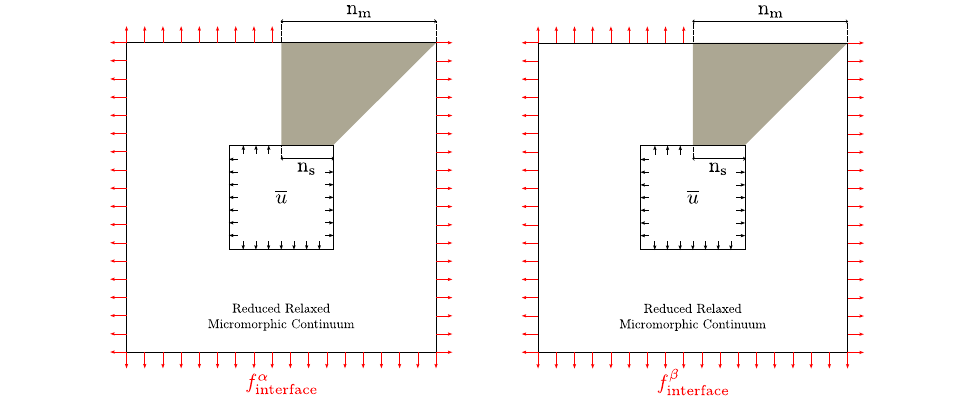}
     \end{subfigure}
        \caption{Reduced relaxed micromorphic formulation of the problem defined in \fig{geometry_free}.}
        \label{fig:rrm_implementation_simple_plate}
\end{figure}

To get a first indication of the influence of the interface forces $f_{\rm interface}$ in the \rrm implementation
of the problem, we start with the special case in which $f_{\rm interface}^{\alpha,\beta}=0$ (which reduces to the simplified conditions of micromorphic free interface given in Equation \eqref{eq:tractions}).
\fig{results_free} shows the comparison of the norm of the displacement field for different frequencies as obtained from:

\begin{itemize}
    \item the microstructured problem defined in Fig. 7 (\textit{first column}) $\beta$ cell’s cut, when setting $\mathrm{n_m}=25$, $\mathrm{n_s}=5$ and the radial expansion displacement $\overline{u}_1 = 1 \times 10^{-12} \frac{0.1x}{x^2+y^2}$[mm], $\overline{u}_2 = 1 \times 10^{-12} \frac{0.1y}{x^2+y^2}$[mm], $\overline{u}_3 = 0$[mm].
    \item the microstructured problem defined in Fig. 7 (\textit{second column}) $\alpha$ cell’s cut, when setting $\mathrm{n_m}=25$, $\mathrm{n_s}=5$ and the radial expansion displacement $\overline{u}_1 = 1 \times 10^{-12} \frac{0.1x}{x^2+y^2}$[mm], $\overline{u}_2 = 1 \times 10^{-12} \frac{0.1y}{x^2+y^2}$[mm], $\overline{u}_3 = 0$[mm].
    \item the reduced relaxed micromorphic implementation of the problem defined in \fig{rrm_implementation_simple_plate} when setting $f_{\rm interface}^{\alpha}=f_{\rm interface}^{\beta}=0$ and the radial expansion displacement $\overline{u}_1 = 1 \times 10^{-12} \frac{0.1x}{x^2+y^2}$[mm], $\overline{u}_2 = 1 \times 10^{-12} \frac{0.1y}{x^2+y^2}$[mm], $\overline{u}_3 = 0$[mm].
    \item As a reference case, we also implement a simulation where the metamaterial’s domain is simulated by using the long-wave limt of our micromorphic continuum \cite{Barbagallo.2017}: a classical Cauchy continuum with tetragonal symmetry whose mechanical properties are given in Table \ref{table:macro_parameters}.
\end{itemize}

\begin{table}[htb]
\centering
\begin{tabular}{||c c c c||} 
 \hline
 $\lambda_{\rm macro}$ & $\mu_{\rm macro}$ & $\mu_{\rm macro}^*$ & $\rho_{\rm macro}$
 \\[1mm]
 [Pa] & [Pa] & [Pa] & [$\rm \frac{kg}{m^3}$]
 \\[1mm]
 \hline\hline
 6.46 $\times 10^7$ & 1.61 $\times 10^9$ & 1.26 $\times 10^6$ & 3975
 \\
 \hline
\end{tabular}
\caption{Parameters of the equivalent tetragonal Cauchy material which is the long-wave limit of the \rrm continuum.}
\label{table:macro_parameters}
\end{table}

We can see from \fig{results_free} that, overall, the influence of the unit cell’s cut on this ``free interface" problem is relatively small, since the main small differences (if any) between the macroscopic displacement fields given in the first two columns are mainly concentrated close to the interface where the displacement $\overline{u}$ is applied.
The overall displacement pattern is thus very similar in both the $\alpha$ and $\beta$ cases.
This leads us to hypothesize that, apart from a small influence of an interface force that would be eventually activated close to the interface where the displacement is applied, we do not expect the activation of an interface force at the
external free interface.
The third column in \fig{results_free} supports our hypothesis, since the reduced relaxed micromorphic model with
vanishing interface forces captures quite well the overall metamaterial’s response for the entire range of considered frequencies.
From the fourth column, we can see that, while an anisotropic equivalent Cauchy continuum can capture to a good extent the
metamaterial’s response at low frequencies, it becomes inadequate when dispersive phenomena start playing a role at higher
frequencies.
We can see that for frequencies which are close or belonging to the band-gap region (1400 and 2000 Hz) some quantitative deviations of the \rrmm from the microstructured solution can be found, even if the overall qualitative behavior remains well captured.
This small deviation lead us to hypothesize that the concentration of the strain across the loading interface due to the fact that local resonances are activated at those frequencies would require in the micromorphic simulation that an interface force is activated close to the loading interface in order to correctly describe the complex mechanisms occurring close to band-gap frequencies, see \fig{non_coherent_1400} and \ref{fig:non_coherent_2000}.\footnotemark

\footnotetext{
To avoid the situation in which a displacement and a force are a applied on the same interface, we slightly modify the simulation setup defined in \fig{rrm_implementation_simple_plate}: instead of assigning the displacement $\overline{u}$ directly on the micromorphic continuum, we put a very thin ring of Cauchy material in perfect contact with the micromorphic continuum. In this way, we can apply the displacement $\overline{u}$ on one side of the Cauchy ring and the interface force as a non-vanishing jump arising at the Cauchy/\rrm interface. Clearly, we checked that the introduction of the Cauchy bar introduces only negligible changes to the solution when considering $\llbracket t \rrbracket = 0$.
}

To understand which type of interface force should be applied close to the interface where the displacement has been assigned, we start by analyzing the Cauchy traction issued via the $\alpha$-cut full-microstructured simulation described in \fig{alpha_beta_free} \textit{(left)}.\footnotemark
\footnotetext{
The Cauchy traction which is here inspected is the one arising at the boundary where the load is applied.
Indeed, in the $\alpha$-cut microstructured simulation, the considered interface is fully solid, so that the traction field can be plotted along the whole surface of interest, see \fig{alpha_beta_free} \textit{(left)}.
Given that the microstructured simulation is implemented using classical elasticity, the traction arising at the $\alpha$-boundary is a Cauchy-like traction.
The plot of the Cauchy traction field along the surface where the displacement is applied, is given in \fig{non_coherent_1400_tractions_microstructured}. 
}
We then explore different expressions of the interface force $f$ in the framework of the \rrm simulation, which are a reasonable interpolation of this Cauchy traction, until the \rrm simulation achieves a good quantitative agreement in terms of displacement field (\fig{non_coherent_1400}).
The interface force which has been needed to achieve a good quantitative agreement at 1400 Hz is shown in \fig{non_coherent_1400_tractions_microstructured}.
We apply the same reasoning for a frequency of 2000 Hz and also for the example of ``metamaterial interface" presented in the next section.\footnotemark
\footnotetext{The choice of the interface forces needed to reproduce the correct macro-displacement field at the targeted frequencies was quite easy to achieve via a simple comparison of \textit{i)} the macroscopic interface force with the corresponding force in the microstructured simulation, \textit{ii)} the effect of the macroscopic interface force on the solution for the displacement field. Further works could be devoted to systematic optimization procedures, once the main features of these macroscopic forces will be integrally unveiled.}

\begin{figure}[H]
     \centering
     \begin{subfigure}[b]{\textwidth}
         \centering
         \includegraphics[width=\textwidth]{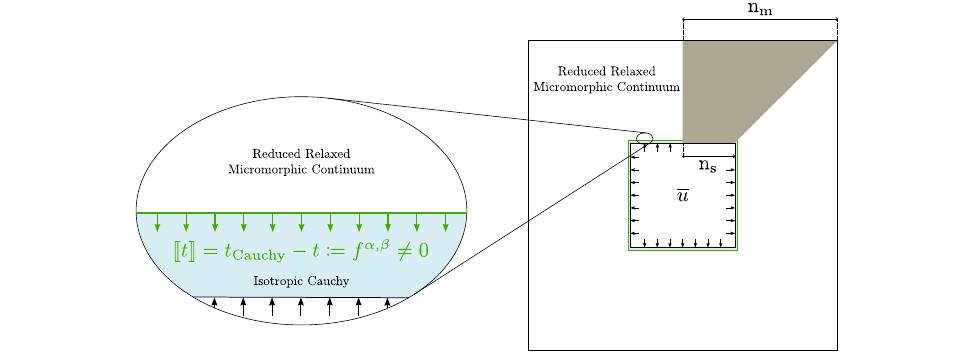}
     \end{subfigure}
        \caption{Reduced relaxed micromorphic implementation of the problem defined in \fig{geometry_free} when a surface force $f^{\alpha,\beta}$ arises close to the loading interface.}
        \label{fig:rrm_implementation_simple_plate_jump}
\end{figure}

\begin{figure}[H]
     \centering
     \begin{subfigure}[b]{\textwidth}
         \centering
         \includegraphics[width=\textwidth]{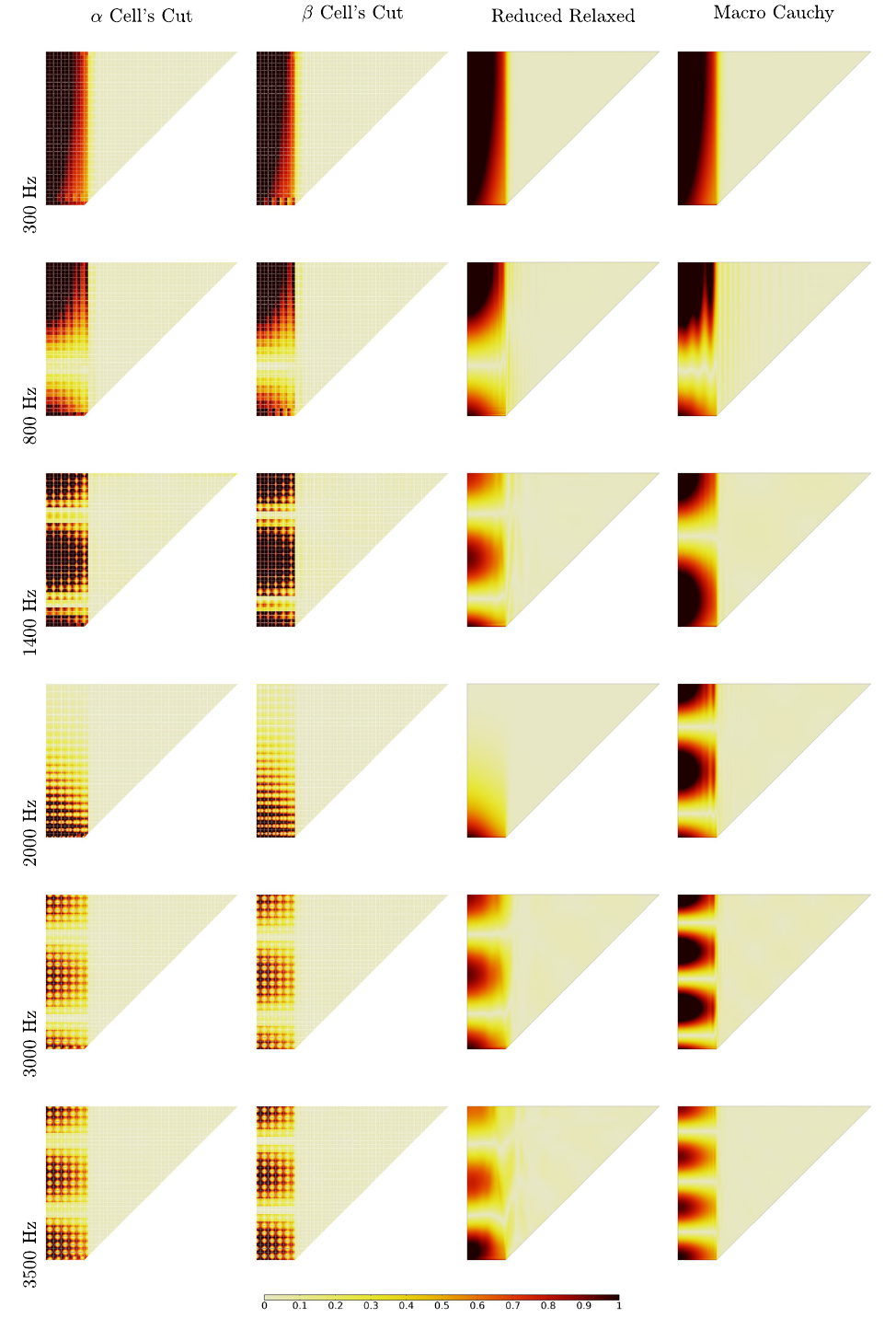}
     \end{subfigure}
        \caption{Results of free interface domain simulations for different frequencies. \textit{(First column)} $\alpha$ unit cell's cut microstructured simulations, \textit{(second column)} $\beta$ unit cell's cuts for different microstructured simulations, \textit{(third column)} \rrm simulations with $f_{\rm interface}=0$, \textit{(fourth column)} tetragonal Cauchy simulations with $f_{\rm interface}=0$.}
        \label{fig:results_free}
\end{figure}

\begin{figure}[H]
     \centering
     \begin{subfigure}[b]{5.789in}
         \centering
         \includegraphics[width=5.789in]{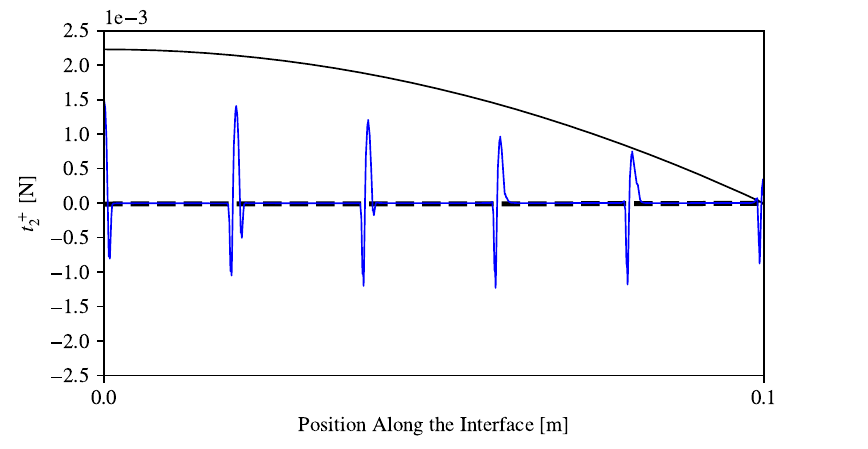}
     \end{subfigure}
        \caption{Plot of the traction at $\omega = 1400$ Hz on the Cauchy side of the metamaterial interface as obtained from \textit{(blue line)} the microstructured simulation, \textit{(dashed line)} the \rrm simulation with no interface force, and \textit{(black line)} the \rrm simulation with the interface force $f_1\!=\!0,\ f_2\!=\!2.25\! \times\! 10^{-3}\dfrac{0.1^2 - x^2}{0.1^2},\ f_3\!=\!0$.}
        \label{fig:non_coherent_1400_tractions_microstructured}
\end{figure}

The quantitative improvement that is obtained at 1400 Hz by using the interface force shown in \fig{non_coherent_1400_tractions_microstructured} is given in \fig{non_coherent_1400}.

\begin{figure}[H]
     \centering
     \begin{subfigure}[b]{\textwidth}
         \centering
         \includegraphics[width=\textwidth]{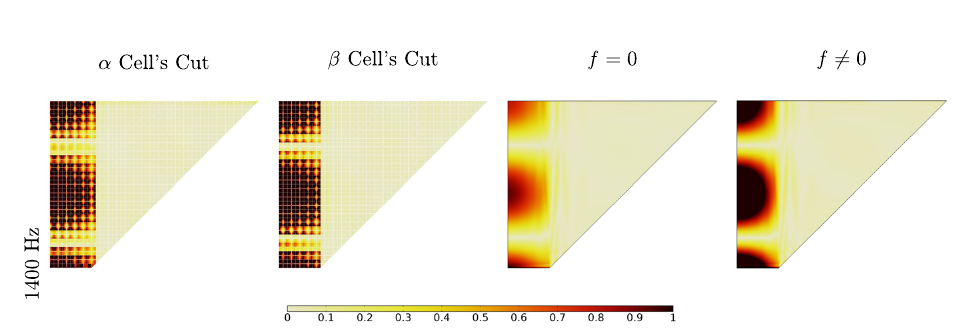}
     \end{subfigure}
        \caption{Comparison at $\omega = 1400$ Hz of free interface \textit{(first column)} $\alpha$ unit cell's cut microstructured simulation, \textit{(second column)} $\beta$ unit cell's cut microstructured simulation, \textit{(third column)} coherent \rrm simulation with $f\!=\!0$, and \textit{(fourth column)} non-coherent \rrm simulation with $f_1\!=\!0,\ f_2\!=\!2.25\! \times\! 10^{-3}\dfrac{0.1^2 - x^2}{0.1^2},\ f_3\!=\!0$. Here, $f_i$ are the components of $f$ of Equation \eqref{eq:interface_non_coherent_conditions_cauchy}.}
        \label{fig:non_coherent_1400}
\end{figure}

\begin{figure}[H]
     \centering
     \begin{subfigure}[b]{5.789in}
         \centering
         \includegraphics[width=5.789in]{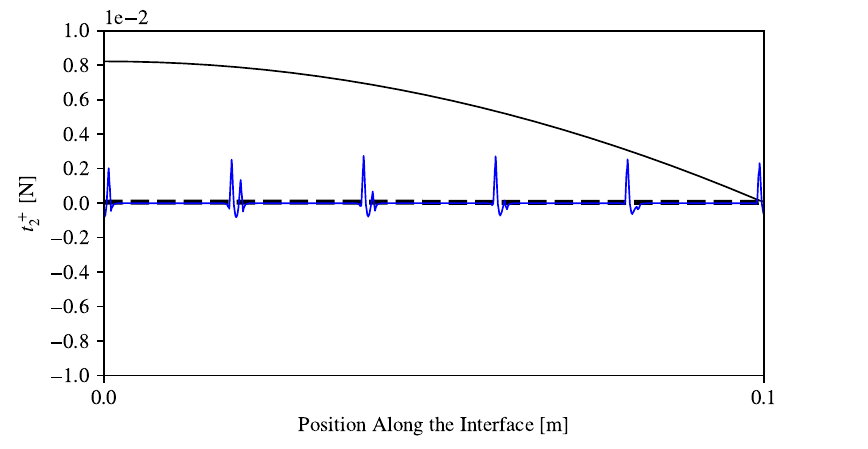}
     \end{subfigure}
        \caption{Plot of the traction at $\omega = 2000$ Hz on the Cauchy side of the metamaterial interface as obtained from \textit{(blue line)} the microstructured simulation, \textit{(dashed line)} the \rrm simulation with no interface force, and \textit{(black line)} the \rrm simulation with the interface force $f_1\!=\!0,\ f_2\!=\!8\! \times\! 10^{-3}\dfrac{0.1^2 - x^2}{0.1^2},\ f_3\!=\!0$.}
        \label{fig:non_coherent_2000_tractions_microstructured}
\end{figure}

\begin{figure}[H]
     \centering
     \begin{subfigure}[b]{\textwidth}
         \centering
         \includegraphics[width=\textwidth]{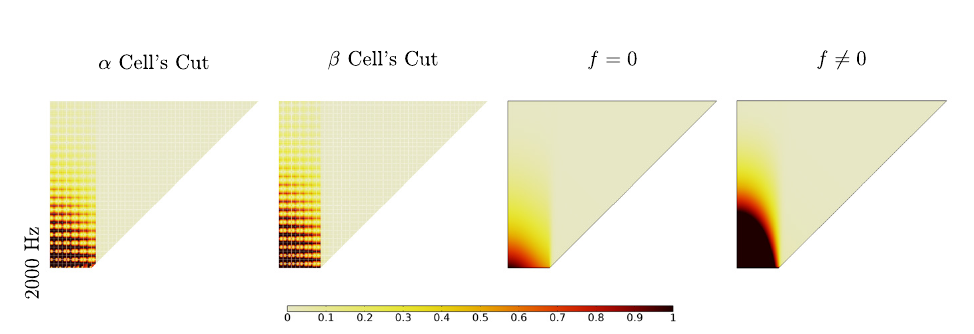}
     \end{subfigure}
        \caption{Comparison at $\omega = 2000$ Hz of free interface \textit{(first column)} $\alpha$ unit cell's cut microstructured simulation, \textit{(second column)} $\beta$ unit cell's cut microstructured simulation, \textit{(third column)} coherent \rrm simulation with $f\!=\!0$, and \textit{(fourth column)} non-coherent \rrm simulation with $f_1\!=\!0,\ f_2\!=\!8\! \times\! 10^{-3}\dfrac{0.1^2 - x^2}{0.1^2},\ f_3\!=\!0$. Here, $f_i$ are the components of $f$ of Equation \eqref{eq:interface_non_coherent_conditions_cauchy}.}
        \label{fig:non_coherent_2000}
\end{figure}

The analogous reasoning leading to the calibration of the interface force at 2000 Hz is shown in \fig{non_coherent_2000_tractions_microstructured}. The quantitative improvement brought by the introduction of such interface force at 2000 Hz is given in \fig{non_coherent_2000}.

\begin{tcolorbox}[colback=Yellow!15!white,colframe=Yellow!50!white,coltitle=black]
Summarizing the main findings of this section, we can say that the ``free interface" problem defined in \fig{geometry_free} is such that:

\begin{itemize}
    \item the macroscopic responses associated to the $\alpha$ and $\beta$ unit cell's cuts are very similar. The only small differences, if any, appear close to the interface where the external load is applied (see \textit{first} and \textit{second column} in \fig{results_free}).
    \item the reduced relaxed micromorphic model, with classical free interface conditions \eqref{eq:tractions}, provides a good qualitative agreement with the microstructured response (\textit{third column}  in \fig{results_free}). However, quantitative deviations can be detected, especially close to the band gap region. These quantitative deviations are corrected when considering suitable interface forces arising close to the loading interface (see \fig{non_coherent_1400} and \ref{fig:non_coherent_2000}).
    \item The main quantitative deviations of the \rrmm with no interface force are found for frequencies close and belonging to the band gap: this is associated to the fact that local resonances concentrated close to the boundary can trigger microstructure-related interface forces.
    \item A quadratic interface force applied in the vicinity of the externally loaded interface was succesful in adjusting the quantititave deviations occuring for frequencies, close and belonging to the band gap region. The intensity of the needed interface force seems to increase with frequency.
    \item The intensity and shape of the \rrm interface force were calibrated using the microstructured traction at the same interface as a reference case.
\end{itemize}
\end{tcolorbox}

\section{Effect of different cell's cuts on a homogeneous-material / metamaterial interface problem}
\label{sec:rmm_5}

In this section, we present a generalization of the problem presented in section \ref{sec:rmm_4}, where the finite-size metamaterial's block shown in \fig{geometry_free} is connected to a homogeneous isotropic material with material parameters $\lambda_{\rm macro}$, $\mu_{\rm macro}$ and $\rho_{\rm macro}$ given in Table \ref{table:macro_parameters}. Analogous simulations could be run by considering the outer material to be titanium, or any other homogeneous material.

\begin{figure}[H]
     \centering
     \begin{subfigure}[b]{\textwidth}
         \centering
         \includegraphics[width=\textwidth]{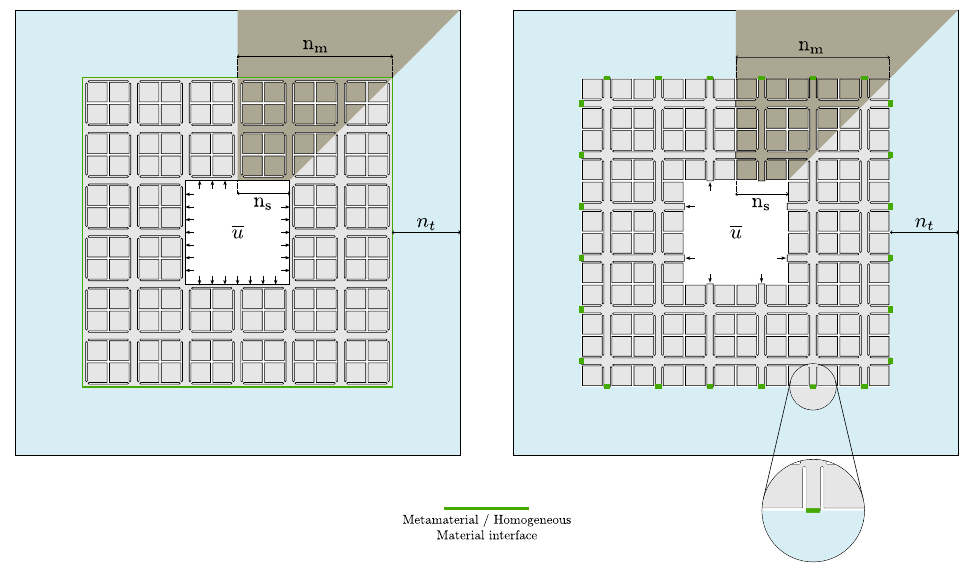}
     \end{subfigure}
        \caption{Full-microstructure setup of the metamaterial/homogeneous material problem for the \textit{(left)} $\alpha$ and \textit{(right)} $\beta$ unit cell's cut.}
        \label{fig:geometry_solid}
\end{figure}

Also in this case, we implement symmetry conditions to reduce computational time (see Appendix \ref{subsec:rmm_7_1}).

\begin{figure}[H]
     \centering
     \begin{subfigure}[b]{\textwidth}
         \centering
         \includegraphics[width=\textwidth]{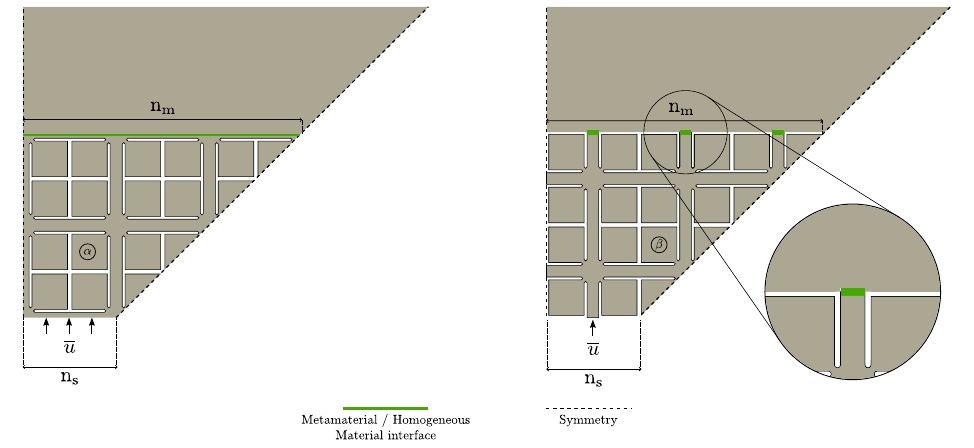}
     \end{subfigure}
        \caption{Symmetrized domain for the metamaterial/homogeneous material interface problem with \textit{(left)} $\alpha$ and \textit{(right)} $\beta$ unit cell's cut.}
        \label{fig:alpha_beta_solid}
\end{figure}

\begin{figure}[H]
     \centering
     \begin{subfigure}[b]{\textwidth}
         \centering
         \includegraphics[width=\textwidth]{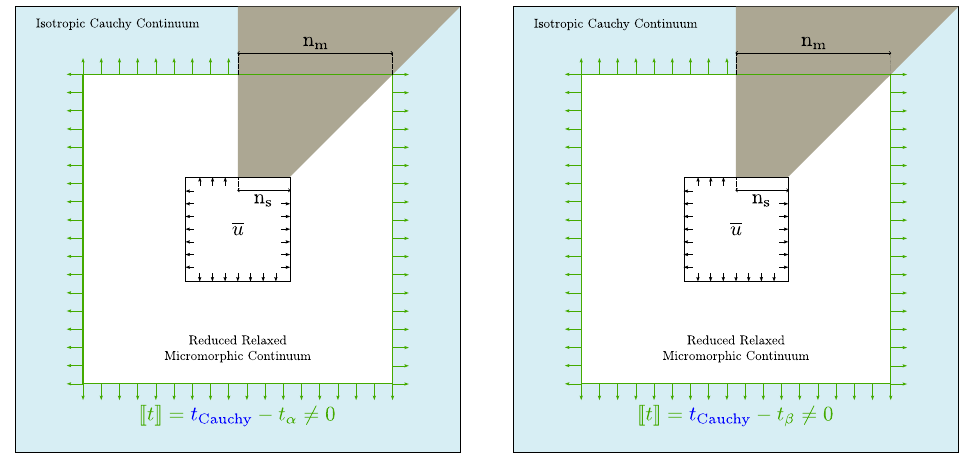}
     \end{subfigure}
        \caption{Reduced relaxed micromorphic formulation of the problem shown in \fig{geometry_solid}. The displacement at the metamaterial's interface is continuous, while the traction is discontinuous and depends on the type of considered unit cell's cut.}
        \label{fig:geometry_rrmm_solid}
\end{figure}

Similarly to what is done in section \ref{sec:rmm_4}, the metamaterial domain of \fig{geometry_solid} is modeled using titanium as the base material of the unit cell, (see Table \ref{table:unit_cell_properties}), whose response is described by classical isotropic linear elasticity.
A perfect connection is supposed to take place at the metamaterial/homogeneous material contact regions, which implies continuity of displacement and of tractions at the solid/solid contacts (green lines in \fig{geometry_solid} or \ref{fig:alpha_beta_solid}).
It is worth to remark that structures of the type presented in \fig{geometry_solid} are easily manufactured via metal etching techniques \cite{Demore.2022}, or also via 3D printing when choosing a polymer as base material.
The \rrm problem corresponding to the one defined in \fig{geometry_solid} is shown in \fig{geometry_rrmm_solid}.
The exterior boundary of the Cauchy medium is left unconstrained in both the microstructured and the \rrm simulations.

\begin{figure}[H]
     \centering
     \begin{subfigure}[b]{\textwidth}
         \centering
         \includegraphics[width=\textwidth]{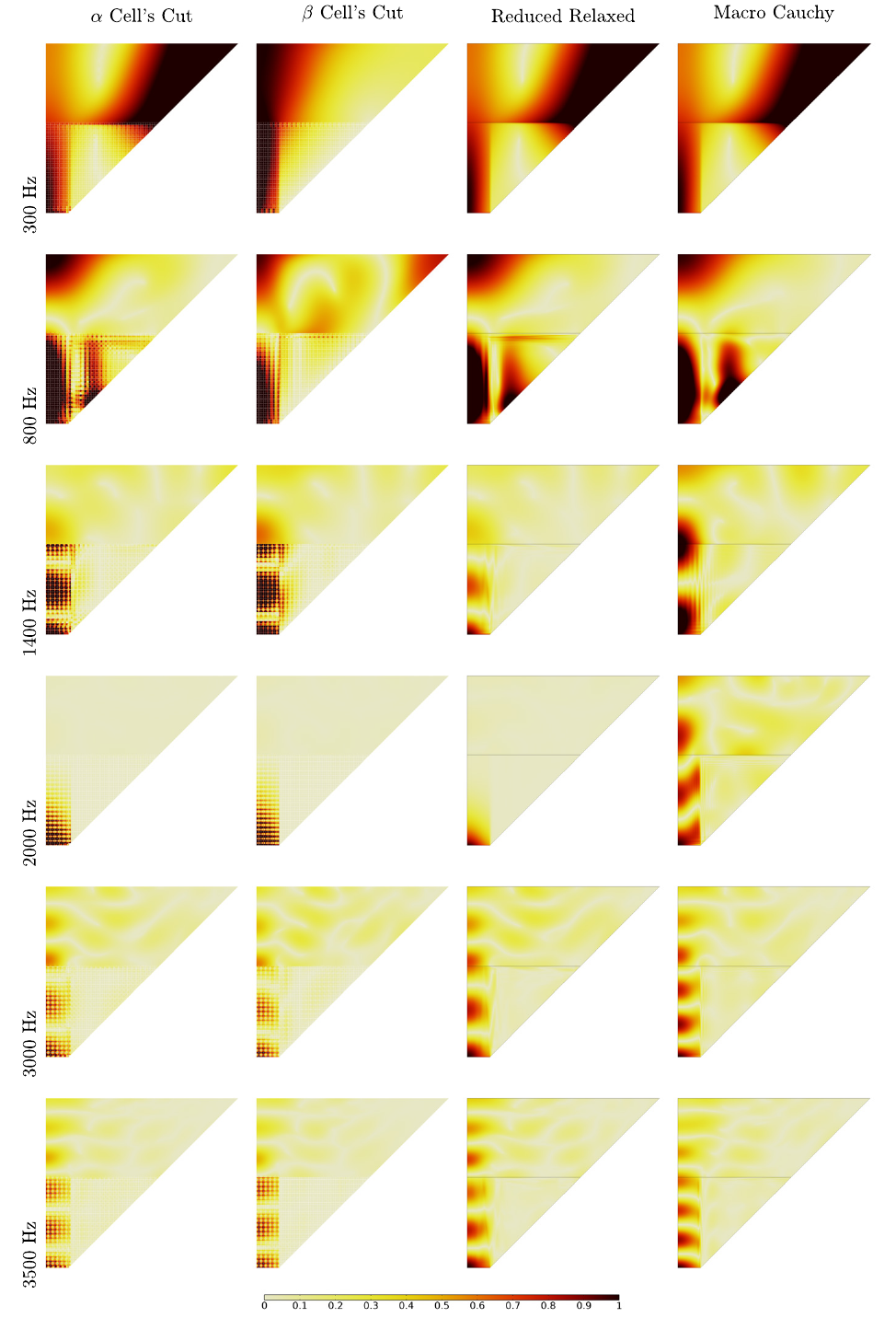}
     \end{subfigure}
        \caption{Results of free interface domain simulations for different frequencies. \textit{(First column)} $\alpha$ unit cell's cut microstructured simulations, \textit{(second column)} $\beta$ unit cell's cuts for different microstructured simulations, \textit{(third column)} \rrm simulations with $\llbracket t \rrbracket=0$, \textit{(fourth column)} tetragonal Cauchy simulations with $\llbracket t \rrbracket = 0$.}
        \label{fig:results_solid}
\end{figure}

\fig{results_solid} shows the solution for the microstructured problem defined in \fig{geometry_solid} for both the $\alpha$ (\textit{first column}) and the $\beta$ (\textit{second column}) cell's cut for different frequencies.
The (\textit{third column}) of \fig{results_solid} shows the particular case of the \rrm problem of \fig{geometry_rrmm_solid}, obtained when setting $\llbracket t \rrbracket=0$.
The \textit{fourth column} shows a reference solution where the metamaterial region is replaced by a Cauchy continuum with tetragonal symmetry, which is the long-wave limit of our micromorphic continuum (see Table \ref{table:macro_parameters} for its elastic coefficients).

By direct inspection of \fig{results_solid}, we can remark that, contrarily to what happens in the ``free interface" case (see \fig{results_free}), the difference between the $\alpha$ and $\beta$ cell's cut is drastic, especially for frequencies below the band-gap region.
This means that the simple change of the interface conditions due to the different $\alpha -$ and $\beta -$type connections to the homogeneous material is responsible for a major change in the bulk solution both in the metamaterial and in the homogeneous material region.
Concerning the \rrm implementation with $\llbracket t \rrbracket=0$ (\textit{third column}), we can see that it gives results very similar to the unit cell's cut $\alpha$ and is thus rather different from the $\beta$ cell solution.
The Cauchy long-wave limit solution with $\llbracket t \rrbracket=0$ is also similar to the $\alpha$ cut solution, even if it starts deviating at relatively low frequencies where only the \rrmm remains accurate.
The fact that the particular \rrm solution with $\llbracket t \rrbracket=0$ is very close to the $\alpha$-type interface is indeed reasonable, since the unit cell's cut $\alpha$ provides a full solid interface in contact with the homogeneous material, which is a situation no too distant from the case of two homogeneous materials in perfect contact (for which the condition $\llbracket t \rrbracket=0$ is known to be the correct one).
On the other hand, the unit cell's cut $\beta$ provides an almost empty interface across which the solid connections are given by thin bars.
It is thus sensible to hypothesize that an interface traction jump $\llbracket t \rrbracket\neq0$ is activated close to the $\beta$ interface.
To prove this, we show in \fig{non_coherent_250}-\ref{fig:non_coherent_2850} the change of solution which is provided when activating a traction jump $\llbracket t \rrbracket \neq 0$.
It can be inferred that when suitably tuning the value of $\llbracket t \rrbracket$, the $\beta -$cell solution can be recovered.
The difference between the $\alpha -$ and $\beta -$cut solutions becomes smaller for frequencies starting around the band-gap region and higher, even if such difference is still present (see \fig{results_solid}).
For frequencies close and belonging to the band gap region, a small correction to the interface force close to the application of the external load, can be given in the \rrm case to improve the slight quantitative difference with respect to the microstructured solution, as already done in the ``free interface" problem in section \ref{sec:rmm_4}.

\begin{figure}[H]
     \centering
     \begin{subfigure}[b]{\textwidth}
         \centering
         \includegraphics[width=\textwidth]{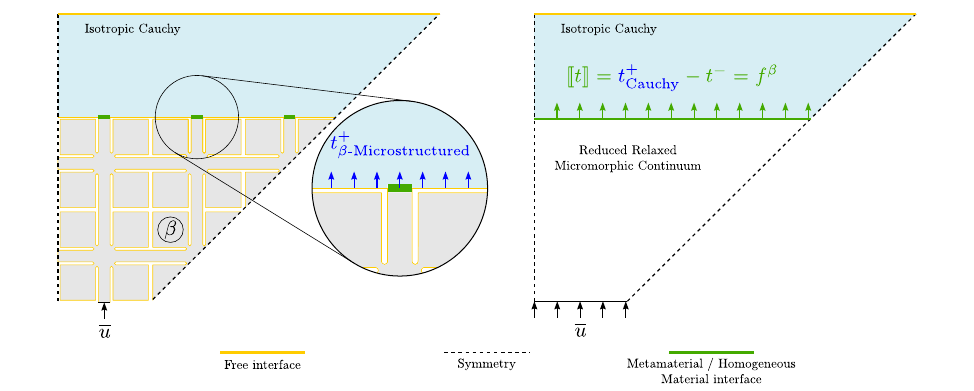}
     \end{subfigure}
        \caption{\textit{(Left)} Representation of the traction $t^+_{\rm \beta-Microstructure}$ arising on the Cauchy side of the microstructured simulation: since the Cauchy material is homogeneous, the interface reported from the Cauchy side is continuous, thus allowing the line-plot of the traction along the interface. \textit{(Right)} Representation of the interface force $f^{\beta}\coloneqq t^+_{\rm Cauchy} - t^-_{\rm \beta-Microstructure}$ at the reduced relaxed micromorphic/Cauchy interface: in order to compare analogous quantities, only $t^+_{\rm \beta-Microstructure}$ and $t^+_{\rm Cauchy}$ will be compared in the following plots.}
        \label{fig:non_coherent_diagram_tractions}
\end{figure}

In order to bring the \rrm solution close to the $\beta$-cut microstructured one, we must look for a suitable form of the interface force triggered at the reduced relaxed micromorphic/homogeneous material interface.
To do so, we proceed similarly to what is done in section \ref{sec:rmm_4} for the ``free interface" case and perform a calibration of the interface force $f^{interface}$ for each frequency.
In particular, we start comparing the traction on the Cauchy side of the interface as arising from the full-microstructured $\beta$-type simulation and the equivalent traction arising from the \rrm simulation when setting $\llbracket t \rrbracket = 0$.
For clarity of exposition, we will denote $t^+_{\rm \beta-Microstructure}$ and $t^+_{\rm Cauchy}$ the traction arising on the Cauchy side of the interface in the microstructured and \rrm simulations, respectively (see \fig{non_coherent_diagram_tractions}) 

\begin{figure}[H]
     \centering
     \begin{subfigure}[b]{5.789in}
         \centering
         \includegraphics[width=5.789in]{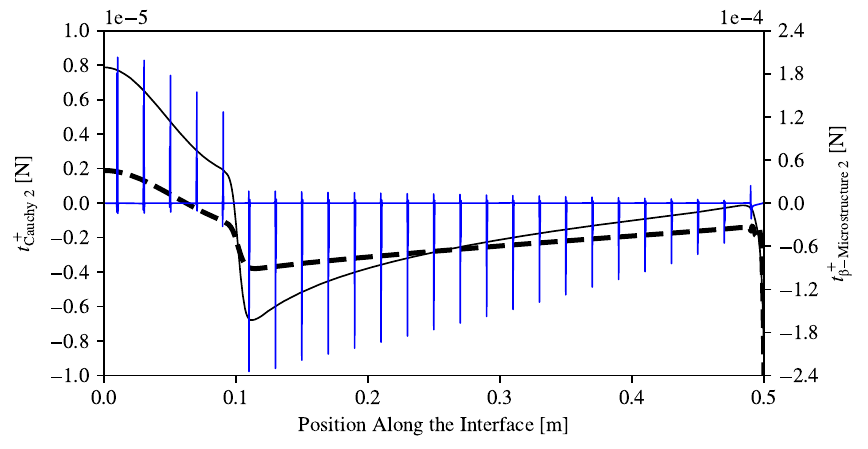}
     \end{subfigure}
        \caption{Plot of the traction at $\omega = 250$ Hz on the Cauchy side of the metamaterial interface as obtained from \textit{(blue line)} the microstructured simulation, \textit{(dashed line)} the \rrm simulation with no interface force, and \textit{(black line)} the \rrm simulation with the interface force $f^{\beta}_1\!=\!0,\ f^{\beta}_2\!=\!-2.2\,t_2^-,\ f^{\beta}_3\!=\!0$.}
        \label{fig:non_coherent_250_tractions_microstructured}
\end{figure}

\begin{figure}[H]
     \centering
     \begin{subfigure}[b]{\textwidth}
         \centering
         \includegraphics[width=\textwidth]{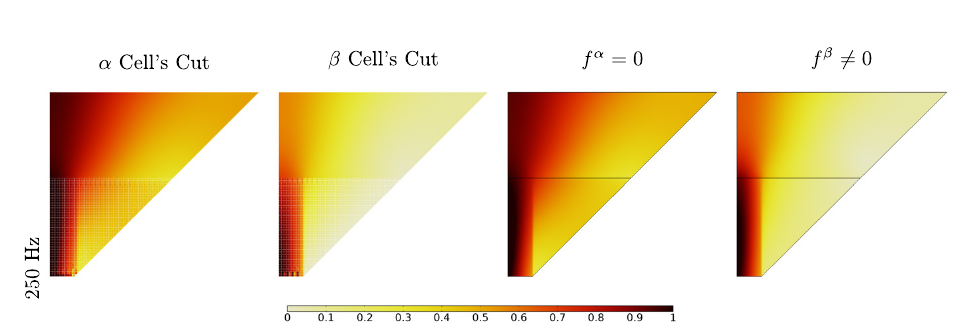}
     \end{subfigure}
        \caption{Comparison at $\omega = 250$ Hz of solid/metamaterial interface
        \textit{(first column)} $\alpha$ unit cell's cut microstructured simulation,
        \textit{(second column)} $\beta$ unit cell's cut microstructured simulation,
        \textit{(third column)} coherent \rrm simulation with $\llbracket t \rrbracket \!=\!f\!=\!f^{\alpha}\!=\!0$, 
        and \textit{(fourth column)} non-coherent \rrm simulation with $\llbracket t \rrbracket \!=\!f\!=\!f^{\beta}\!\neq\!0$ and $f^{\beta}_1\!=\!0,\ f^{\beta}_2\!=\!-2.2\,t_2^-,\ f^{\beta}_3\!=\!0$.
        Here, $t^-_i$ and $f_i$ are the components of the micromorphic traction $t^-$ and of the force $f$ of Equation \eqref{eq:interface_non_coherent_conditions_cauchy}.}
        \label{fig:non_coherent_250}
\end{figure}

\begin{figure}[H]
     \centering
     \begin{subfigure}[b]{5.789in}
         \centering
         \includegraphics[width=5.789in]{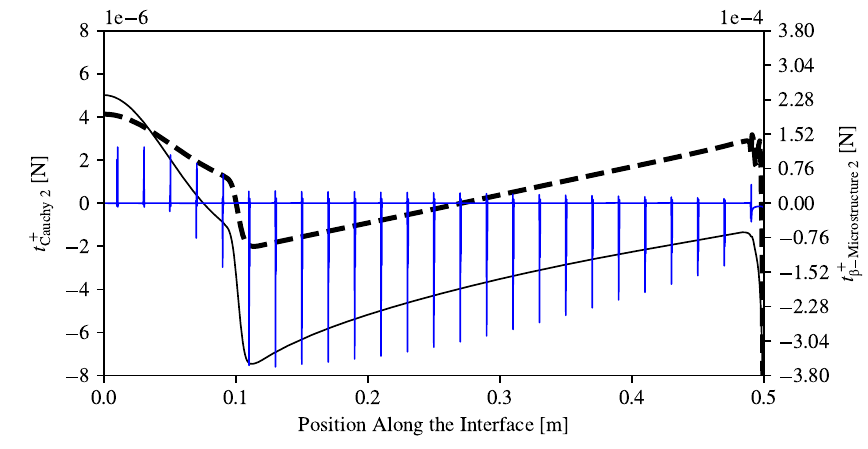}
     \end{subfigure}
        \caption{Plot of the traction at $\omega = 300$ Hz on the Cauchy side of the metamaterial interface as obtained from \textit{(blue line)} the microstructured simulation, \textit{(dashed line)} the \rrm simulation with no interface force, and \textit{(black line)} the \rrm simulation with the interface force $f^{\beta}_1\!=\!0,\ f^{\beta}_2\!=\!-1.6\,t_2^-,\ f^{\beta}_3\!=\!0$.}
        \label{fig:non_coherent_300_tractions_microstructured}
\end{figure}

\begin{figure}[H]
     \centering
     \begin{subfigure}[b]{\textwidth}
         \centering
         \includegraphics[width=\textwidth]{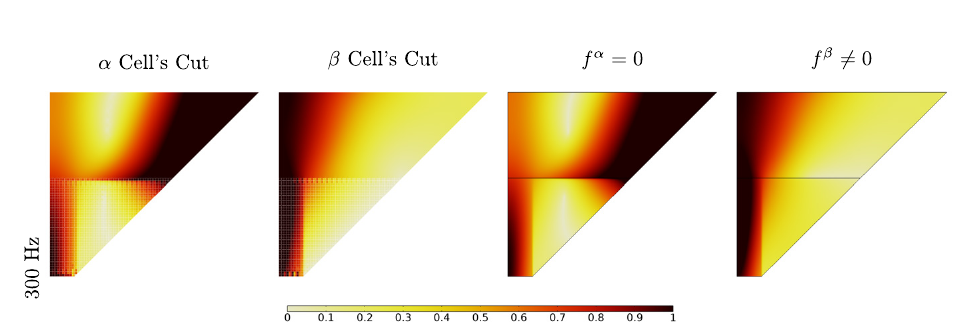}
     \end{subfigure}
        \caption{Comparison at $\omega = 300$ Hz of solid/metamaterial interface
        \textit{(first column)} $\alpha$ unit cell's cut microstructured simulation,
        \textit{(second column)} $\beta$ unit cell's cut microstructured simulation,
        \textit{(third column)} coherent \rrm simulation with $\llbracket t \rrbracket \!=\!f\!=\!f^{\alpha}\!=\!0$, and
        \textit{(fourth column)} non-coherent \rrm simulation with $\llbracket t \rrbracket \!=\!f\!=\!f^{\beta}\!\neq\!0$ and $f^{\beta}_1\!=\!0,\ f^{\beta}_2\!=\!-1.6\,t_2^-,\ f^{\beta}_3\!=\!0$.
        Here, $t^-_i$ and $f_i$ are the components of the micromorphic traction $t^-$ and of the force $f$ of Equation \eqref{eq:interface_non_coherent_conditions_cauchy}.}
        \label{fig:non_coherent_300}
\end{figure}

\begin{figure}[H]
     \centering
     \begin{subfigure}[b]{5.789in}
         \centering
         \includegraphics[width=5.789in]{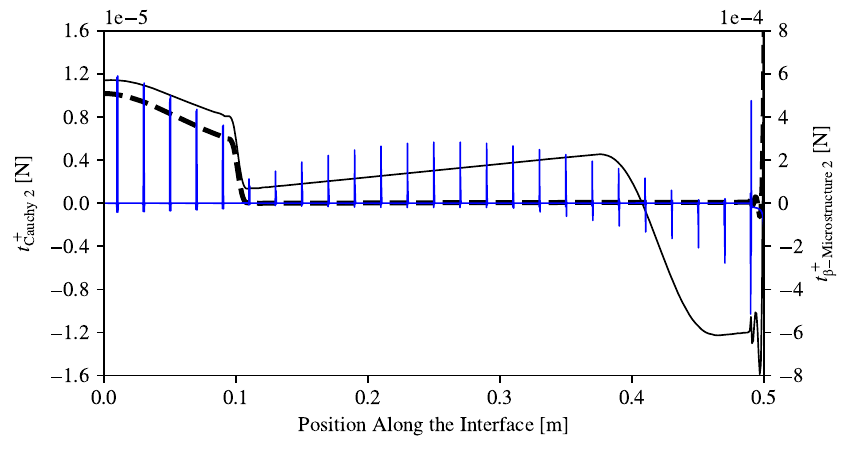}
     \end{subfigure}
        \caption{Plot of the traction at $\omega = 450$ Hz on the Cauchy side of the metamaterial interface as obtained from \textit{(blue line)} the microstructured simulation, \textit{(dashed line)} the \rrm simulation with no interface force, and \textit{(black line)} the \rrm simulation with the interface force $f^{\beta}_1\!=\!0,\ f^{\beta}_2\!=\!\dfrac{6\! \times\! 10^{-6}}{0.5}x - 1.8\! \times\! 10^{-5}\mathrm{H}\left( x - 0.42\right),\ f^{\beta}_3\!=\!0$.}
        \label{fig:non_coherent_450_tractions_microstructured}
\end{figure}

\begin{figure}[H]
     \centering
     \begin{subfigure}[b]{\textwidth}
         \centering
         \includegraphics[width=\textwidth]{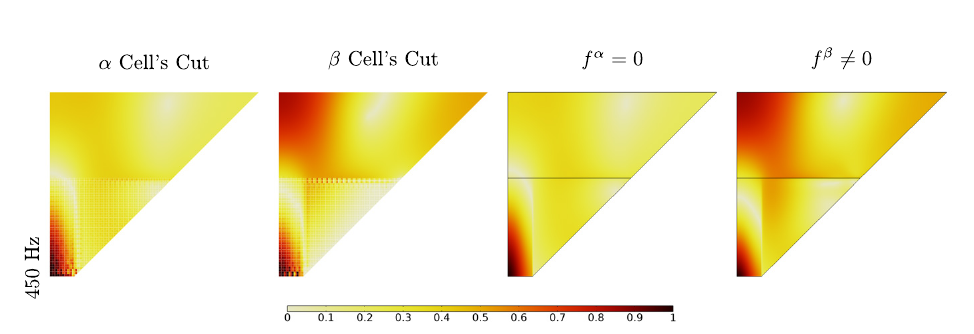}
     \end{subfigure}
        \caption{Comparison at $\omega = 450$ Hz of solid/metamaterial interface
        \textit{(first column)} $\alpha$ unit cell's cut microstructured simulation,
        \textit{(second column)} $\beta$ unit cell's cut microstructured simulation,
        \textit{(third column)} coherent \rrm simulation with $\llbracket t \rrbracket \!=\!f\!=\!f^{\alpha}\!=\!0$, and
        \textit{(fourth column)} non-coherent \rrm simulation with $\llbracket t \rrbracket \!=\!f\!=\!f^{\beta}\!\neq\!0$ and $f^{\beta}_1\!=\!0,\ f^{\beta}_2\!=\!\dfrac{6\! \times\! 10^{-6}}{0.5}x - 1.8\! \times\! 10^{-5}\mathrm{H}\left( x - 0.42\right),\ f^{\beta}_3\!=\!0$.
        Here, $f_i$ are the components of the force $f$ of Equation \eqref{eq:interface_non_coherent_conditions_cauchy}, and H is the Heaviside step function with a transition zone of $0.1 \mathrm{\ m}$.}
        \label{fig:non_coherent_450}
\end{figure}

\begin{figure}[H]
     \centering
     \begin{subfigure}[b]{5.789in}
         \centering
         \includegraphics[width=5.789in]{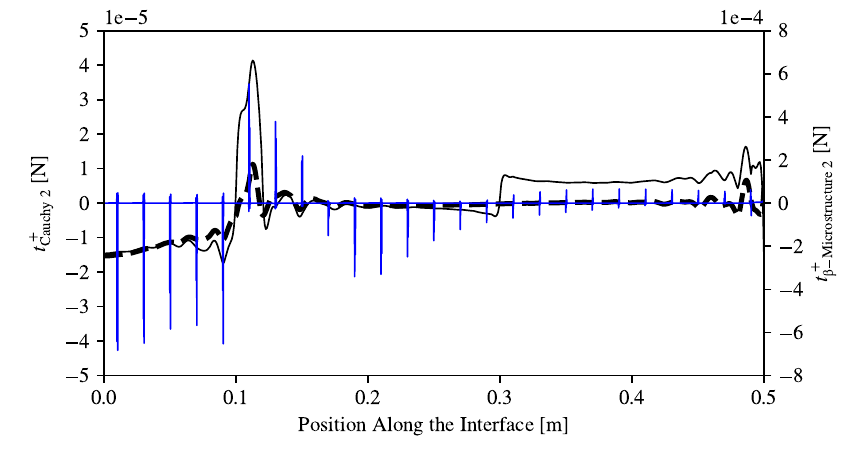}
     \end{subfigure}
        \caption{Plot of the traction at $\omega = 1050$ Hz on the Cauchy side of the metamaterial interface as obtained from \textit{(blue line)} the microstructured simulation, \textit{(dashed line)} the \rrm simulation with no interface force, and \textit{(black line)} the \rrm simulation with the interface force $f^{\beta}_1\!=\!0,\ f^{\beta}_2\!=\!3\! \times\! 10^{-6}\mathrm{H}\left( 0.1 - x \right) + 3\! \times\! 10^{-5}\left( \mathrm{H}\left( x - 0.1 \right) - \mathrm{H}\left( 0.12 - x \right) \right) + 1.2\! \times\! 10^{-5}\mathrm{H}\left( x - 0.3 \right),\ f^{\beta}_3\!=\!0$.}
        \label{fig:non_coherent_1050_tractions_microstructured}
\end{figure}

\begin{figure}[H]
     \centering
     \begin{subfigure}[b]{\textwidth}
         \centering
         \includegraphics[width=\textwidth]{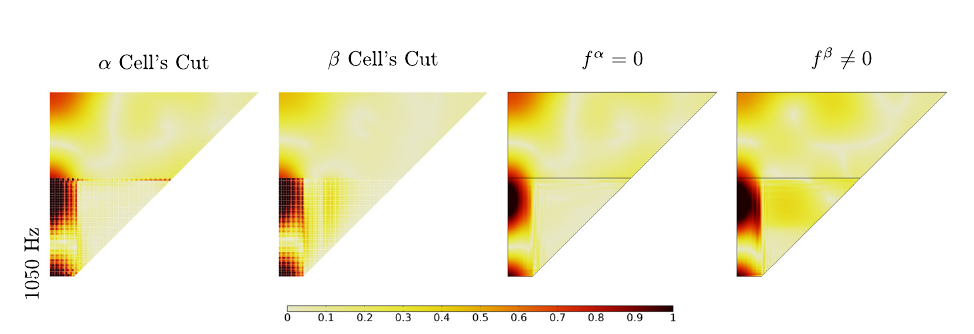}
     \end{subfigure}
        \caption{Comparison at $\omega = 1050$ Hz of solid/metamaterial interface
        \textit{(first column)} $\alpha$ unit cell's cut microstructured simulation,
        \textit{(second column)} $\beta$ unit cell's cut microstructured simulation,
        \textit{(third column)} coherent \rrm simulation with $\llbracket t \rrbracket \!=\!f\!=\!f^{\alpha}\!=\!0$, and
        \textit{(fourth column)} non-coherent \rrm simulation with $\llbracket t \rrbracket \!=\!f\!=\!f^{\beta}\!\neq\!0$ and $f^{\beta}_1\!=\!0,\ f^{\beta}_2\!=\!3\! \times\! 10^{-6}\mathrm{H}\left( 0.1 - x \right) + 3\! \times\! 10^{-5}\left( \mathrm{H}\left( x - 0.1 \right) - \mathrm{H}\left( 0.12 - x \right) \right) + 1.2\! \times\! 10^{-5}\mathrm{H}\left( x - 0.3 \right),\ f^{\beta}_3\!=\!0$.
        Here, $f_i$ are the components of the force $f$ of Equation \eqref{eq:interface_non_coherent_conditions_cauchy}, and H is the Heaviside step function with a transition zone of $0.005 \mathrm{\ m}$.}
        \label{fig:non_coherent_1050}
\end{figure}

\begin{figure}[H]
     \centering
     \begin{subfigure}[b]{5.789in}
         \centering
         \includegraphics[width=5.789in]{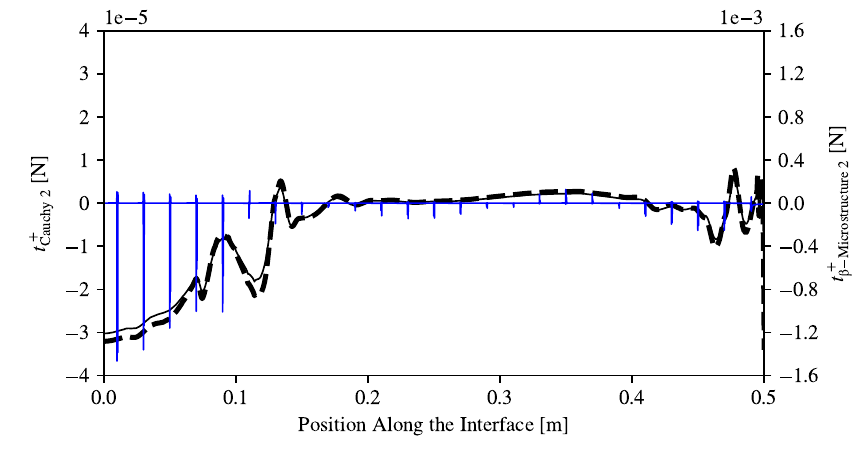}
     \end{subfigure}
        \caption{Plot of the traction at $\omega = 2850$ Hz on the Cauchy side of the $\beta$ unit cell metamaterial interface as obtained from \textit{(blue line)} the microstructured simulation, \textit{(dashed line)} the \rrm simulation with no interface force, and \textit{(black line)} the \rrm simulation with the interface force $f^{\beta}_1\!=\!0,\ f^{\beta}_2\!=\!0.2\,t_2^-,\ f^{\beta}_3\!=\!0$.}
        \label{fig:non_coherent_2850_tractions_beta_microstructured}
\end{figure}

\begin{figure}[H]
     \centering
     \begin{subfigure}[b]{5.789in}
         \centering
         \includegraphics[width=5.789in]{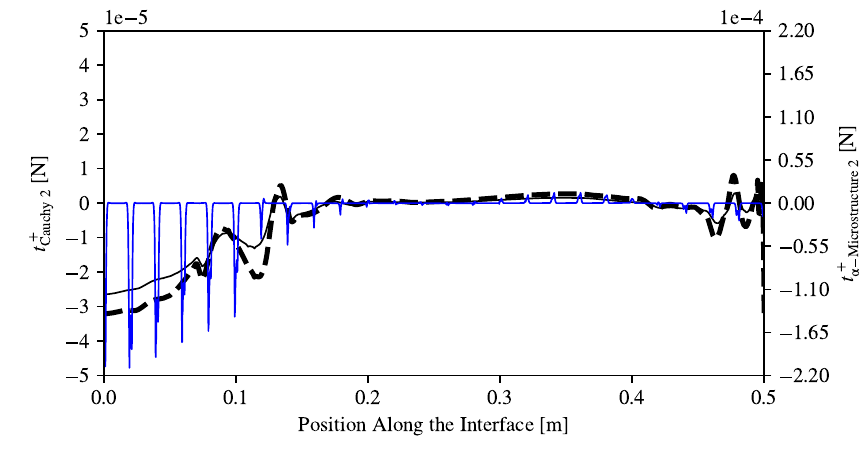}
     \end{subfigure}
        \caption{Plot of the traction at $\omega = 2850$ Hz on the Cauchy side of the $\alpha$ unit cell metamaterial interface as obtained from \textit{(blue line)} the microstructured simulation, \textit{(dashed line)} the \rrm simulation with no interface force, and \textit{(black line)} the \rrm simulation with the interface force $f^{\alpha}_1\!=\!0,\ f^{\alpha}_2\!=\!0.45\,t_2^-,\ f^{\alpha}_3\!=\!0$.}
        \label{fig:non_coherent_2850_tractions_alpha_microstructured}
\end{figure}

\begin{figure}[H]
     \centering
     \begin{subfigure}[b]{\textwidth}
         \centering
         \includegraphics[width=\textwidth]{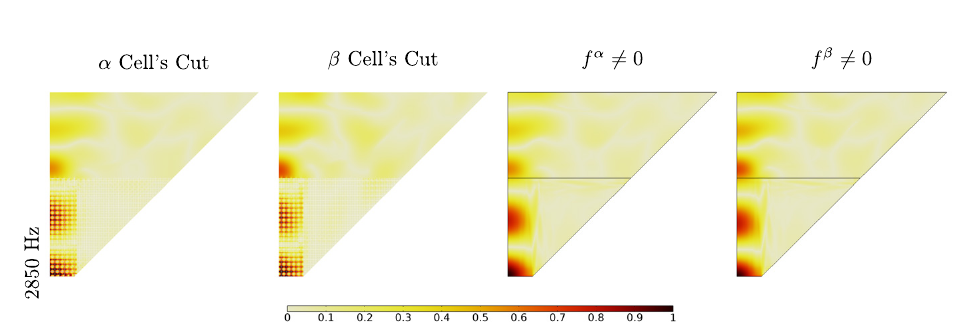}
     \end{subfigure}
        \caption{Comparison at $\omega = 2850$ Hz of solid/metamaterial interface
        \textit{(first column)} $\alpha$ unit cell's cut microstructured simulation,
        \textit{(second column)} $\beta$ unit cell's cut microstructured simulation,
        \textit{(third column)} non-coherent \rrm simulation with $\llbracket t \rrbracket \!=\!f\!=\!f^{\alpha}\!\neq\!0$ and $f^{\alpha}_1\!=\!0,\ f^{\alpha}_2\!=\!0.45\,t_2^-,\ f^{\alpha}_3\!=\!0$, and
        \textit{(fourth column)} non-coherent \rrm simulation with $\llbracket t \rrbracket \!=\!f\!=\!f^{\beta}\!\neq\!0$ and $f^{\beta}_1\!=\!0,\ f^{\beta}_2\!=\!0.2\,t_2^-,\ f^{\beta}_3\!=\!0$.
        Here, $t^-_i$ and $f_i$ are the components of the micromorphic traction $t^-$ and of the force $f$ of Equation \eqref{eq:interface_non_coherent_conditions_cauchy}.}
        \label{fig:non_coherent_2850}
\end{figure}

The calibration procedure for the interface force at $\omega=250$ Hz is shown in \fig{non_coherent_250_tractions_microstructured}.
As mentioned above, we start by inspecting the behavior of the Cauchy traction of the \rrm simulation with $\llbracket t \rrbracket = 0$ \textit{(dashed lined)} and comparing it with the trend of the Cauchy traction issued from the full-microstructured simulation.
We then make an ansatz on the interface force to be such that $f^{\rm interface}=\alpha t_{\rm micromorphic}$ and we see which is the effect of varying the $\alpha$ parameter on the Cauchy traction.
This simplified ansatz on the form of $f^{\rm interface}$ was sufficient to retrieve the correct solution for some frequencies.
For other frequencies the ansatz on $f^{\rm interface}$ had to be more complex and it is indicated in the captions of the corresponding figures.

For the frequency $\omega=2850$ Hz, an interface force was needed to adjust both the $\alpha$ and $\beta$ cuts: this might be due to the fact that for higher frequencies the wavelength becomes comparable to the cell's size. Two different calibration procedures were thus needed as shown in \fig{non_coherent_2850_tractions_beta_microstructured} and \ref{fig:non_coherent_2850_tractions_alpha_microstructured}.

As shown for the ``free interface" case, small quantitative deviations of the \rrm solutions from the microstructured ones can be found for frequencies close or inside the band gap (see e.g. $\omega = 2000$ Hz in \fig{results_solid}).
As shown before for the ``free interface case", those small deviations can be easily adjusted considering a quadratic interface force close to the application of the external load.
Since the results are analogous to the ones presented before, we do not present specific pictures here and we refer to \fig{non_coherent_2000} for possible reference.

\begin{tcolorbox}[colback=Yellow!15!white,colframe=Yellow!50!white,coltitle=black]
The analysis of \fig{non_coherent_250}-\ref{fig:non_coherent_2850} brings us to important conclusions, significantly advancing the state of the art concerning the macroscopic modeling of mechanical metamaterials.
We summarize the main findings as follows:

\begin{itemize}
    \item The metamaterial interface which is connected to the homogeneous material through a full-solid connection ($\alpha$ unit cell type) is well described by the \rrmm with interface conditions $\llbracket u \rrbracket = 0$ and $\llbracket t \rrbracket = 0$.
    \item The metamaterial interface which is connected to the homogeneous material with thin bars and empty-space elsewhere ($\beta$ unit cell type) can no longer be described by the boundary conditions $\llbracket u \rrbracket = 0$ and $\llbracket t \rrbracket = 0$, when considering the \rrm formulation of the problem. In this case, the macroscopic \rrm/Cauchy interface must be considered to be an elastic interface: $\llbracket u \rrbracket = 0$ and $\llbracket t \rrbracket \neq 0$.
    \item The difference between the $\alpha -$ and $\beta -$type solutions is more pronounced in the frequency region which goes from approximately 2000 Hz to the lower band-gap limit (approx. 1700 Hz). Given that such differences were not present in the ``free interface" problem (\fig{results_free}), this is most likely due to the fact that the wavelengths in this frequency range have a characteristic size comparable with the length of the metamaterial/homogeneous material interface (1 m), while for frequencies higher than 2000 Hz the wavelenghts become small with respect to the interface. This suggests that the considered interface indeed acts as an ``obstacle" for the traveling waves which behaves differently in the $\alpha$ and $\beta$ case. In other words, we can say that, for the considered frequencies, such a macroscopic interface can be seen as a ``material interface" with its own elastic properties that are conferred to the interface itself from the characteristics of the underlying microstructure.
    \item For very low frequencies (up to 300-400 Hz), the metamaterial response can be caught rather well by a macroscopic Cauchy model with tetragonal symmetry. This means that at low frequencies the anisotropy alone is sufficient to account for the macroscopic heterogeneity driven by the underlying microstructure. Using ``classical" interface conditions or ``elastic interface" conditions allows to discriminate between the $\alpha$ and the $\beta$ solution also in the simplified case of Cauchy elasticity, when the frequency remains small enough to avoid the activation of dispersive phenomena.
    \item For frequencies higher than 400 Hz, dispersive phenomena start taking place due to a more complex interaction of the wave with the underlying microstructure, so that a \rrmm is necessary to catch this complex response. As said before, the \rrmm can discriminate between the $\alpha$ and $\beta$ solutions by switching from the condition $\llbracket t \rrbracket=0$ to the condition $\llbracket t \rrbracket\neq0$.
    \item Considering adequate interface forces at the reduced relaxed micromorphic/homogeneous material interface can account for the large differences in the solution between the $\alpha$ and $\beta$ cut for frequencies lower than the band-gap. 
    \item Small quantitative deviations of the \rrm solutions from the microstructured solution close and inside the band-gap region, can once again be re-adjusted by considering the presence of an interface force close to the application of the external load.
\end{itemize}
\end{tcolorbox}

\section{Conclusions and perspective}
\label{sec:rmm_6}

In this paper, we showed for the first time that the modeling of mechanical metamaterials at the homogenized scale necessarily requires the introduction of the concept of ``non-coherent" interfaces when considering metamaterial specimens of finite-size.
In particular, we proved that the introduction of interface forces which are able to discriminate between two different unit cell's cuts, is of paramount importance if one wants to use macroscopic (homogenized) models to describe the response of metamaterial's blocks of finite-size.
The need of introducing such interface forces has been clearly explained through the presentation of two different examples in which interface forces are defined in the context of \rrm elasticity.
The results presented in this paper are a fundamental milestone for the understanding of the way in which metamaterials should be modeled when considering finite-size blocks at the engineering scale.
Given the novelty of the obtained results, the present paper opens new perspectives for further studies which will be oriented to understand the effect on interface forces of:

\begin{itemize}
    \item the macroscopic geometry of the bulk domains and of the interfaces and intensity,
    \item the type and intensity of the applied load,
    \item the size of the metmaterial's specimen, and
    \item the considered frequency.
\end{itemize}

\section{Annexes}
\label{sec:rmm_7}

\subsection{Symmetry Conditions of the \rrm Continuum}
\label{subsec:rmm_7_1}
We follow the approach of \cite{Demore.2022} to define the symmetry conditions of the \rrm continuum. Let us apply Curie's Symmetry Principle\cite{Curie.1894} on $u \in \mathbb{R}^{3}$ the macroscopic displacement field and on $P \in \mathbb{R}^{3\times 3}$ the non-symmetric micro-distortion field, respect to a symmetry plane $\mathcal{N}$ of normal $n$:

\begin{align}
	\left\{
	\arraycolsep=1pt\def\arraystretch{1}
	\begin{array}{ccc}
		u \left( x^* \right) 
		& = 
		& u^* \left( x \right) 
        \\
		P \left( x^* \right) 
		& = 
		& P^* \left( x \right)
	\end{array}
	\right.
	,
	\label{eq:curie_sym}
\end{align}

\noindent
where $u^*,P^*$ are the symmetrics of $u,P$. We also define the corresponding orthonormal bases $\left\{ t^*_1,t^*_2,n^*\right\}$ and $\left\{ t_1,t_2,n\right\}$ using vectors tangent and normal to $\mathcal{N}$, such that 

\begin{equation}
    t^*_1=t_1\ ,\ t^*_2=t_2\ ,\ \mathrm{and}\ n^*=-n\ .
    \label{eq:bases_transformation}
\end{equation}

We write the expansion of the $u$ macroscopic displacement and $P$ micro-distorsion fields relative to their corresponding basis:

\begin{align}
    u =     & u_1 t_1 + u_2 t_2 + u_3 n \ ,\label{eq:u}\\
    P =     & P_{11} t_1 \otimes t_1 + P_{12} t_1 \otimes t_2 + P_{13} t_1 \otimes n \notag\\
            & + P_{21} t_2 \otimes t_1 + P_{22} t_2 \otimes t_2 + P_{23} t_2 \otimes n \notag\\
            & + P_{31} n \otimes t_1 + P_{32} n \otimes t_2 + P_{33} n \otimes n \ ,\label{eq:P}\\
            &\mathrm{and} \notag\\
    u^* =   & u_1 t^*_1 + u_2 t^*_2 + u_3 n^* \ ,\label{eq:u_star}\\
    P^* =   & P_{11} t^*_1 \otimes t^*_1 + P_{12} t^*_1 \otimes t^*_2 + P_{13} t^*_1 \otimes n^* \notag\\
            & + P_{21} t^*_2 \otimes t^*_1 + P_{22} t^*_2 \otimes t^*_2 + P_{23} t^*_2 \otimes n^* \notag\\
            & + P_{31} n^* \otimes t^*_1 + P_{32} n^* \otimes t^*_2 + P_{33} n^* \otimes n^* \ ,\label{eq:P_star}
\end{align}

\noindent
where the scalars $u_i$ and $P_{ij}$ are the unique components of $u$ and $P$ relative to both bases, and in both bases the same. If we use the equivalencies of Equation \eqref{eq:bases_transformation}, we can rewrite \eqref{eq:u_star} and \eqref{eq:P_star} like:

\begin{align}
    u^* =   & u_1 t_1 + u_2 t_2 - u_3 n \ ,\label{eq:u_star_n}\\
    P^* =   & P_{11} t_1 \otimes t_1 + P_{12} t_1 \otimes t_2 - P_{13} t_1 \otimes n \notag\\
            & + P_{21} t_2 \otimes t_1 + P_{22} t_2 \otimes t_2 - P_{23} t_2 \otimes n \notag\\
            & - P_{31} n \otimes t_1 - P_{32} n \otimes t_2 + P_{33} n \otimes n \ ,\label{eq:P_star_n}
\end{align}

Similarly, if we define the points $x$ and $x^*$ as $x=x_0 + \epsilon n$ and $x^*=x_0 + \epsilon n^*$, where $x_0 \in \mathcal{N}$ and $\epsilon \in \mathbb{R}$, we can say that $x^*=x_0 - \epsilon n$. Next, if we write the symmetry conditions \eqref{eq:curie_sym} using the expansions of Equations \eqref{eq:u},\eqref{eq:P} and \eqref{eq:u_star_n},\eqref{eq:P_star_n}, and group them in terms of their components, we reach:

\begin{align}
	\left\{
	\arraycolsep=1pt\def\arraystretch{1}
	\begin{array}{ccc}
		u_1 \left( x_0 - \epsilon n \right) t_1   &=& u_1 \left( x_0 + \epsilon n \right) t_1\\
		u_2 \left( x_0 - \epsilon n \right) t_2   &=& u_2 \left( x_0 + \epsilon n \right) t_2\\
        u_3 \left( x_0 - \epsilon n \right) n     &=& -u_3 \left( x_0 + \epsilon n \right) n
	\end{array}
	\right.
    \quad
	,
    \quad
    \left\{
	\arraycolsep=1pt\def\arraystretch{1}
	\begin{array}{ccc}
		P_{11} \left( x_0 - \epsilon n \right) t_1 \otimes t_1  &=& P_{11} \left( x_0 + \epsilon n \right) t_1 \otimes t_1\\
		P_{12} \left( x_0 - \epsilon n \right) t_1 \otimes t_2  &=& P_{12} \left( x_0 + \epsilon n \right) t_1 \otimes t_2\\
        P_{13} \left( x_0 - \epsilon n \right) t_1 \otimes n    &=& -P_{13} \left( x_0 + \epsilon n \right) t_1 \otimes n\\
        P_{21} \left( x_0 - \epsilon n \right) t_2 \otimes t_1  &=& P_{21} \left( x_0 + \epsilon n \right) t_2 \otimes t_1\\
		P_{22} \left( x_0 - \epsilon n \right) t_2 \otimes t_2  &=& P_{22} \left( x_0 + \epsilon n \right) t_2 \otimes t_2\\
        P_{23} \left( x_0 - \epsilon n \right) t_2 \otimes n    &=& -P_{23} \left( x_0 + \epsilon n \right) t_2 \otimes n\\
        P_{31} \left( x_0 - \epsilon n \right) n \otimes t_1    &=& -P_{31} \left( x_0 + \epsilon n \right) n \otimes t_1\\
		P_{32} \left( x_0 - \epsilon n \right) n \otimes t_2    &=& -P_{32} \left( x_0 + \epsilon n \right) n \otimes t_2\\
        P_{33} \left( x_0 - \epsilon n \right) n \otimes n      &=& P_{33} \left( x_0 + \epsilon n \right) n \otimes n
	\end{array}
	\right.
    \quad .
	\label{eq:curie_sym_pre_limit}
\end{align}

These symmetry conditions allow to find the values of $u$ and $P$ at any point in space with respect to the symmetry plane, if their value is known on the opposite side of the symmetry plane. In the limit of $\epsilon \to 0$, Equation \eqref{eq:curie_sym_pre_limit} becomes:

\begin{align}
	\left\{
	\arraycolsep=1pt\def\arraystretch{1}
	\begin{array}{ccc}
		u_1 \left( x_0 \right) t_1   &=& u_1 \left( x_0 \right) t_1\\
		u_2 \left( x_0 \right) t_2   &=& u_2 \left( x_0 \right) t_2\\
        u_3 \left( x_0 \right) n     &=& -u_3 \left( x_0 \right) n
	\end{array}
	\right.
    \quad
	,
    \quad
    \left\{
	\arraycolsep=1pt\def\arraystretch{1}
	\begin{array}{ccc}
		P_{11} \left( x_0 \right) t_1 \otimes t_1  &=& P_{11} \left( x_0 \right) t_1 \otimes t_1\\
		P_{12} \left( x_0 \right) t_1 \otimes t_2  &=& P_{12} \left( x_0 \right) t_1 \otimes t_2\\
        P_{13} \left( x_0 \right) t_1 \otimes n    &=& -P_{13} \left( x_0 \right) t_1 \otimes n\\
        P_{21} \left( x_0 \right) t_2 \otimes t_1  &=& P_{21} \left( x_0 \right) t_2 \otimes t_1\\
		P_{22} \left( x_0 \right) t_2 \otimes t_2  &=& P_{22} \left( x_0 \right) t_2 \otimes t_2\\
        P_{23} \left( x_0 \right) t_2 \otimes n    &=& -P_{23} \left( x_0 \right) t_2 \otimes n\\
        P_{31} \left( x_0 \right) n \otimes t_1    &=& -P_{31} \left( x_0 \right) n \otimes t_1\\
		P_{32} \left( x_0 \right) n \otimes t_2    &=& -P_{32} \left( x_0 \right) n \otimes t_2\\
        P_{33} \left( x_0 \right) n \otimes n      &=& P_{33} \left( x_0 \right) n \otimes n
	\end{array}
	\right.
    \quad .
	\label{eq:curie_sym_post_limit}
\end{align}

We find that on the symmetry plane $\mathcal{N}$: $u_3=P_{13}=P_{23}=P_{31}=P_{32}=0$. Thus, the following conditions must be satisfied on $\mathcal{N}$:

\begin{align}
	\left\{
	\arraycolsep=1pt\def\arraystretch{1}
	\begin{array}{ccc}
		\left\langle u,n \right\rangle &=& 0\\
		\left\langle P,t_1 \otimes n \right\rangle &=& 0\\
        \left\langle P,t_2 \otimes n \right\rangle &=& 0\\
        \left\langle P,n \otimes t_1 \right\rangle &=& 0\\
        \left\langle P,n \otimes t_2 \right\rangle &=& 0
	\end{array}
	\right.
    \quad .
	\label{eq:curie_sym_on_plane}
\end{align}

For a plane strain problem, the displacement field has $u_3=0$ and the microdistorsion tensor has $P_{3i}=0$. Thus, for a vertical symmetry plane: $t_1=(0,0,1),t_2=(0,1,0),n=(1,0,0)$, with $t_1$ and $t_2$ arbitrary tangents, the symmetry conditions \eqref{eq:curie_sym_on_plane} reduce to:

\begin{align}
	\left\{
	\arraycolsep=1pt\def\arraystretch{1}
	\begin{array}{cccc}
		\left\langle u,n \right\rangle                &= & u_1    &= 0\\
		\left\langle P,t_1 \otimes n \right\rangle    &= & 0      &= 0\\
        \left\langle P,t_2 \otimes n \right\rangle    &= & P_{21} &= 0\\
        \left\langle P,n \otimes t_1 \right\rangle    &= & 0      &= 0\\
        \left\langle P,n \otimes t_2 \right\rangle    &= & P_{12} &= 0
	\end{array}
	\right.
    \quad .
	\label{eq:curie_sym_at_0}
\end{align}

Equivalently, for a symmetry plane at 45\textdegree: $t_1=(0,0,1),t_2=(\dfrac{1}{\sqrt{2}},\dfrac{1}{\sqrt{2}},0),n=(-\dfrac{1}{\sqrt{2}},\dfrac{1}{\sqrt{2}},0)$, with $t_1$ and $t_2$ arbitrary tangents, the symmetry conditions \eqref{eq:curie_sym_on_plane} reduce to:

\begin{align}
	\left\{
	\arraycolsep=1pt\def\arraystretch{1}
	\begin{array}{cccc}
		\left\langle u,n \right\rangle                &= & -u_1 + u_2 &= 0\\
		\left\langle P,t_1 \otimes n \right\rangle    &= & 0      &= 0\\
        \left\langle P,t_2 \otimes n \right\rangle    &= & -P_{11}+P_{12}-P_{21}+P_{22} &= 0\\
        \left\langle P,n \otimes t_1 \right\rangle    &= & 0      &= 0\\
        \left\langle P,n \otimes t_2 \right\rangle    &= & -P_{11}-P_{12}+P_{21}+P_{22} &= 0
	\end{array}
	\right.
    \quad .
	\label{eq:curie_sym_at_45}
\end{align}

{\scriptsize
	\paragraph{{\scriptsize Acknowledgements.}}
	Angela Madeo, Leonardo A. Perez Ramirez, Félix Erel-Demore, and Gianluca Rizzi acknowledge support from the European Commission through the funding of the ERC Consolidator Grant META-LEGO, N$^\circ$ 101001759.
}



\begingroup
\setstretch{1}
\setlength\bibitemsep{3pt}
\printbibliography
\endgroup

\end{document}